\documentclass[sigplan,10pt,nonacm]{acmart}
\usepackage{paralist}
\usepackage{xspace}
\usepackage{comment}
\usepackage{epsfig}
\PassOptionsToPackage{hyphens}{url}
\usepackage{color}

\usepackage{ragged2e}
\usepackage{ifthen}
\usepackage{amsmath}
\usepackage{subcaption}
\usepackage{tikz}
\usepackage{caption}
\usepackage{multicol}
\usepackage{xcolor}
\usepackage[moderate,title,sections,leadingfraction=1,charwidthfraction=1]{savetrees}

\renewcommand{\paragraph}[1]{\vspace{0.2ex}\noindent{\bf #1} }

\newboolean{publicversion}
\setboolean{publicversion}{false}

\ifthenelse{\boolean{publicversion}}{
  \newcommand{\grumbler}[2]{}
}{
  \newcommand{\grumbler}[2]{\textcolor{blue}{\bf #1: #2}}
}

\definecolor{heraldBlue}{rgb}{0.0,0.0,0.8}
\definecolor{heraldRed}{rgb}{0.8,0.0,0.0}
\definecolor{heraldGray}{rgb}{0.2,0.2,0.5}
\definecolor{heraldGreen}{rgb}{0.0,0.4,0.0}
\definecolor{heraldPink}{rgb}{0.8,0.1,0.6}

\newcommand*{\eg}{e.g.\@\xspace}

\makeatletter
\newcommand*{\etc}{%
    \@ifnextchar{.}%
        {etc}%
        {etc.\@\xspace}%
}
\makeatother

\newcommand{\sys}[0]{Assise\xspace}
\newcommand{\libfs}[0]{LibFS\xspace}
\newcommand{\kernfs}[0]{SharedFS\xspace}

\newcommand{\Sys}[0]{\sys}
\newcommand{\ccnvm}[0]{CC-NVM\xspace}

\newcommand*\circled[1]{\tikz[baseline=(char.base)]{
            \node[shape=circle,fill,inner sep=.2pt] (char)
            {\textcolor{white}{\small \\#1}};}}

\renewcommand\footnotetextcopyrightpermission[1]{}
\title{\Large \bf \Sys: Performance and Availability via NVM
  Colocation in a Distributed File System}
\author{Thomas E. Anderson}
\affiliation{University of Washington}
\author{Marco Canini}
\affiliation{KAUST}
\author{Jongyul Kim}
\authornotemark[2]
\affiliation{KAIST}
\author{Dejan Kosti\'c}
\affiliation{KTH Royal Institute of Technology}
\author{Youngjin Kwon}
\affiliation{KAIST}
\author{Simon Peter}
\affiliation{UT Austin}
\author{Waleed Reda}
\authornote{Lead student author.}
\affiliation{Universit\'{e} catholique de Louvain}
\affiliation{KTH Royal Institute of Technology}
\author{Henry N. Schuh}
\authornote{Co-student authors.}
\affiliation{University of Washington}
\author{Emmett Witchel}
\affiliation{UT Austin}
\date{}

\begin{document}

\settopmatter{printacmref=false, printfolios=true}
\maketitle



\subsection*{Abstract}
The adoption of very low latency persistent memory modules (PMMs)
upends the long-established model of disaggregated file system access.
%
%
Instead, by colocating computation and PMM storage, we can provide
applications much higher I/O performance, sub-second application
failover, and strong consistency. To demonstrate this, we built the
\sys distributed file system, based on a persistent, replicated
coherence protocol for managing a set of server-colocated PMMs as a
fast, \emph{crash-recoverable} cache between applications and slower
disaggregated storage, such as SSDs. Unlike disaggregated file
systems, \sys maximizes locality for all file IO by carrying out IO on
colocated PMM whenever possible and minimizes coherence overhead by
maintaining consistency at IO operation granularity, rather than at
fixed block sizes.

We compare \sys to Ceph/Bluestore, NFS, and Octopus on a cluster with
Intel Optane DC PMMs and SSDs for common cloud applications and
benchmarks, such as LevelDB, Postfix, and FileBench. We find that \sys
improves write latency up to 22$\times$, throughput up to 56$\times$,
fail-over time up to 103$\times$, and scales up to 6$\times$ better
than its counterparts, while providing stronger consistency
semantics. \Sys promises to beat the MinuteSort world record by
1.5$\times$.

\section{Introduction}

Byte-addressable non-volatile memory (NVM), such as Intel's Optane DC persistent
memory module (PMM)~\cite{optanedc}, is now commercially available as main
memory. NVM provides high-capacity persistent memory with near-DRAM performance at
much lower cost and energy use. The promise of NVM as a low-cost main memory add-on
is driving the adoption of node-colocated NVM at
scale~\cite{baidu_3dxpoint,google_3dxpoint,oracle_3dxpoint}. With remote direct
memory access (RDMA), non-local NVM can be accessed across the network without CPU
overhead, raising interest in NVM also for high-performance distributed storage.

A common paradigm in distributed file systems (like Amazon EFS~\cite{efs},
NFS~\cite{nfsv4}, Ceph~\cite{ceph}, Colossus/GFS~\cite{gfs} and even NVM-aware
re-designs, like Octopus~\cite{octopus} and Orion~\cite{orion}) is to separate
storage servers from clients. In this disaggregated design, files are stored by
servers on machines physically separated from clients running applications. Client
main memory is treated as a volatile block cache managed by the client's OS kernel.
Storage disaggregation simplifies resource pooling by physically separating
application from storage concerns with simple, server-managed data consistency
mechanisms.


Disaggregation's simplicity comes at a cost, which becomes apparent as we move from
SSD/HDD to NVM storage.  First, in steady state, applications are slowed by the
system call overhead to access kernel-level client caches, and (on cache misses) by
the need for multiple network round trips to consult disaggregated meta-data servers
and then to access the actual data.  Second, on failure, disaggregated file systems
must rebuild data and metadata caches of failed clients from scratch, causing long
recovery times to reestablish application-level service and with high network
utilization during recovery.  Third, managing client caches at fixed page-block
granularity amplifies the small IO operations typical of many of today's distributed
applications and may cause additional cache coherence overhead when IO is larger than
the block size. These costs inhibit the full utilization of NVM performance and have
led some within the storage community to advocate for a complete redesign of the file
system API~\cite{stonebraker,snia,levelhash,recipe}.

We present \sys, a distributed file system designed to maximize the use of {\em
  colocated} NVM without requiring a new API for high performance. \Sys unleashes the
performance of NVM via pervasive and persistent caching in process-local,
socket-local, node-local, and network-local NVM. \sys accelerates POSIX file IO by
orders of magnitude by leveraging colocated NVM without kernel involvement, block
amplification, or unnecessary coherence overheads. \Sys provides near-instantaneous
application fail-over onto a \emph{cache replica} that mirrors an application's local
file system cache in the replica's colocated NVM. \Sys reduces node recovery time by
orders of magnitude by locally recovering NVM caches with strong consistency
semantics. Finally, \sys leverages network-local NVM via \emph{reserve replicas} that
provide colocated NVM as a next-level cache with lower latency than disaggregated
SSDs. In cascaded failure scenarios, reserve replicas become cache replicas to
preserve near-instantaneous fail-over.

To enable these properties, we designed and built to our knowledge the first crash
consistent distributed file system cache coherence layer for replicated NVM (\ccnvm).
\ccnvm serves cached file system state in \sys with strong consistency guarantees and
locality. \ccnvm provides crash consistency with prefix semantics~\cite{salus} by
enforcing write order to local NVM via logging and to cross-socket and remote NVM by
leveraging the write ordering of DMA and RDMA, respectively. \ccnvm provides
linearizability for all IO operations via leases~\cite{gray89leases} that can be
delegated among nodes, sockets, and processes for direct local management of file
system state. \ccnvm consistently chain-replicates~\cite{chain_replication} all file
system updates to a configurable set of replicas for availability.

Using \ccnvm, \sys achieves the following goals:

\begin{compactitem}[\labelitemi]
\item \textbf{Simple programming model.} \Sys supports unmodified applications using
  the familiar POSIX API with strong linearizability and crash consistency
  semantics~\cite{salus}.

\item \textbf{Scalability.} Unlike NVM-aware distributed file systems that are
  limited to rack-scale~\cite{orion,legoos}, \sys provides strong consistency but
  remains scalable using dynamic delegation of leases to nodes and processes;
  node-local and socket-local sharing uses \ccnvm for consistency without network or
  cross-socket communication.

\item \textbf{Low IO tail latency.} To efficiently support applications with low tail
  latency requirements, \sys allows kernel-bypass access to authorized local and
  remote NVM areas. To optionally lower write latency with replicated persistence,
  \sys provides an optimistic mode using asynchronous chain replication with prefix
  crash consistency semantics.

\item \textbf{High availability.}
  Distributed applications require high storage system uptime.
  \Sys provides near-instantaneous application fail-over to a configurable number of
  replicas, while minimizing the time to restore the system replication factor after
  a failure.

\item \textbf{Efficient bandwidth use.} The high bandwidth provided by NVM means that
  communication can be a throughput bottleneck
  (cf. Table~\ref{tab:storage_hierarchy}). \Sys minimizes communication by using
  update logs for consistency, allowing it to eliminate redundant
  writes~\cite{strata} and coherence protocol overhead.
\end{compactitem}



\noindent We make the following contributions.

\begin{compactitem}[\labelitemi]
\item We present the design (\S\ref{sec:design}) and implementation
  (\S\ref{sec:impl}) of \sys, a distributed file system that fully utilizes NVM by
  persistent caching in colocated NVM as a primary design principle. \Sys is the
  first distributed file system to recover the file system cache for fast fail-over
  and to locally synchronize reads and writes to file system state.

\item We present \ccnvm (\S\ref{sec:ccnvm}), the first persistent and available
  distributed cache coherence layer. \ccnvm provides locality for data and meta-data
  access, replicates for availability, and provides linearizability and crash
  consistency with prefix semantics for all (shared) file system IO.

\item We quantify the performance benefits of NVM colocation versus disaggregation
  for distributed file systems (\S\ref{sec:eval}).  We compare \sys's steady-state
  and fail-over behavior to RDMA-accelerated versions of Ceph with
  Bluestore~\cite{bluestore} and NFS, as well as Octopus~\cite{octopus}, a
  distributed file system designed for RDMA and NVM, using common cloud applications
  and benchmarks, such as LevelDB, Postfix, MinuteSort, and FileBench.
\end{compactitem}



\noindent 
Our evaluation shows that \sys provides up to 22$\times$ lower write latency and up
to 56$\times$ higher throughput than NFS and Ceph/Bluestore. \Sys also outperforms
Octopus by up to an order of magnitude for these workloads. \Sys scales better than
Ceph, providing 6$\times$ higher mail delivery throughput with Postfix at scale. \Sys
is more available than Ceph, returning a recovering LevelDB store to full performance
up to 103$\times$ faster. Showing that strong consistency with the familiar POSIX API
and high performance are not mutually exclusive, \sys finishes a local external sort
3\% faster than a hand-tuned implementation using processor loads and stores to
memory mapped NVM. Finally, \sys finishes the MinuteSort distributed sorting
benchmark up to 2.2$\times$ faster than a parallel NFS installation. Scaling \sys's
result to the cluster size of the current MinuteSort world record~\cite{minutesort},
we can estimate that \sys sorts 1.5$\times$ more data per minute than the record
holder.

\Sys still supports disaggregated storage where it makes sense. \Sys can
automatically migrate cold data that does not fit in NVM to slower, disaggregated
cold storage devices, such as SSDs and HDDs. To do so, \sys's implementation builds
on Strata~\cite{strata} as its node-local store. We describe both in the appendix
(\S\ref{app:iopaths})). We will release \sys as open source.

\section{Background}\label{sec:background}

\begin{table}
\footnotesize
  \centering
  \begin{tabular}{lrrr}
    Memory & R/W Latency & Seq. R/W GB/s & \$/GB \\
    \hline
    DDR4 DRAM   & 82 ns     &  107 / 80  & 35.16~\cite{optanedc_price} \\
    NVM (local) & 175 / 94 ns     & 32 / 11.2 & 4.51~\cite{optanedc_price} \\
    NVM-NUMA & 230 ns     & 4.8 / 7.4 & - \\
    NVM-kernel & 0.6 / 1 $\mu$s & - & - \\
    NVM-RDMA & 3 / 8 $\mu$s   & 3.8      & - \\
    SSD (local)  & 10 $\mu$s  & 2.4 / 2.0   & 0.32~\cite{intel_dc_p4610} \\
  \end{tabular}
  \caption{Memory \& storage price/performance as of May 2020.\vspace{-2ex}}
  \label{tab:storage_hierarchy}
\end{table}


Distributed applications have diverse workloads, with IO granularities
large and small~\cite{ursa}, different sharding patterns, and
consistency requirements. All demand high availability and
scalability. Supporting all of these properties simultaneously has
been the focus of decades of distributed storage
research~\cite{ceph,octopus,orion,wafl,zfs,xfs,nfsv4}. Before NVM,
trade-offs had to be made. For example, by favoring large transfers
ahead of small IO, performance ahead of crash consistency, or common
case performance ahead of fast recovery, leading to the common
disaggregated file system design. We argue that with the advent of
fast NVM, these trade-offs need to be re-evaluated.

The opportunity posed by NVM is two-fold:

\paragraph{Cost/performance.}\label{sec:nvm_trends}
Table~\ref{tab:storage_hierarchy} shows measured access latency,
bandwidth, and cost for modern server memory and storage technologies,
including Optane DC PMM (measurement details in \S\ref{sec:eval}). We
can see that local NVM access latency and bandwidth are near-DRAM, up
to two orders of magnitude better than SSD. At the same time, NVM's
per-GB cost is 13\% that of DRAM. NVM's unique characteristics allow
it to be used as the top layer in the storage hierarchy, as well as
the bottom layer in a server's memory hierarchy.

\paragraph{Fast recovery.} Persistent local storage with near-DRAM
performance can provide a \emph{recoverable} cache for hot file system
data that can persist across reboots. The vast majority of system
failures are due to software crashes that simply require
rebooting~\cite{google_study,ibm_study,hennessy_patterson}. Caching
hot file system data in NVM allows for quick recovery from these
common failures.

For these reasons, data center operators are deploying NVM at
scale~\cite{google_3dxpoint,baidu_3dxpoint,oracle_3dxpoint}. However,
to fully utilize its potential, we have to efficiently use colocated
NVM. NVM accessed via RDMA (NVM-RDMA), via loads and stores to another
CPU socket (NVM-NUMA), or via system calls on the same socket
(NVM-kernel) can be an order of magnitude slower in terms of both
latency and bandwidth.

\subsection{Related Work}\label{sec:related}

We survey the existing work in distributed storage and highlight why
it is unlikely to fully utilize the storage system performance offered
by the integration of colocated NVM.

\vspace{-.2ex}
\paragraph{Block and object stores,} such as Amazon's EBS~\cite{ebs},
S3~\cite{s3}, and Ursa~\cite{ursa}, provide a new API to a multi-layer
storage hierarchy that can provide cheap, fault-tolerant access to
vast amounts of data.  However, block stores have a minimum IO
granularity (16KB for EBS) and IO smaller than the block size suffers
performance degradation from write
amplification~\cite{raju17sosp-pebblesdb,ursa}. To illustrate, Dropbox
uses Amazon S3 for data blocks, but keeps small metadata in DRAM for
fast access, backed by an SSD~\cite{dropbox}. Apache
Crail~\cite{crail} and Blizzard~\cite{blizzard} provide file system
APIs on top of block stores, but both focus on parallel throughput of
large data streams, rather than small IO.

To realize the performance benefits of NVM for all IO, we need to
abandon fixed block sizes and instead persist and track IO at its
original operation granularity. Hence, \sys leverages logging to
persist and enforce the consistency of writes at their original
granularity in NVM. A similar model is realized in the
RAMcloud~\cite{ramcloud} key-value store. RAMcloud maintains data in
DRAM for performance, using SSDs for asynchronous
persistence. However, the capacity limits of DRAM mean that many
RAMcloud operations still involve the network, and because DRAM state
cannot be recovered after a crash, it is vulnerable to cascading node
failures. Even after single node failures, state must be restored from
remote nodes. RAMcloud requires a full-bisection bandwidth network for
fast recovery. \Sys leverages colocated NVM for recovery and does not
require full-bisection bandwidth or asynchronous backup storage.

\vspace{-.2ex}
\paragraph{Disaggregated file systems,} like Ceph~\cite{ceph}, use
distributed hashing over nodes to provide scalable file service for
cloud applications.  However, network and system call latency harms
file IO latency as shown in Table~\ref{tab:storage_hierarchy}. High
throughput via parallel network access to disaggregated NVM is
similarly surpassed by the higher bandwidth of colocated NVM.

To combat network overheads, several disaggregated file systems have
been built~\cite{octopus,orion} or
retrofitted~\cite{accelio,hdfs_nvm,nfsv4} to use
RDMA. Octopus~\cite{octopus} and Orion~\cite{orion} are prominent
redesigns that use RDMA for high performance access to NVM. Still,
neither leverages kernel-bypass for low-latency IO (Octopus uses FUSE,
Orion runs in the kernel) and both are disaggregated designs. Like
Ceph, Octopus uses distributed hashing to place files on nodes
(Octopus does not replicate). Orion can store data locally (for
\emph{internal clients}), but uses a disaggregated metadata server.
Both systems perform multiple remote operations per file IO in the
common case to update file and/or metadata, increasing IO latency.

Network latency and limited bandwidth increase file system operation
latency, reduce throughput, and limit scalability. Due to update
contention at the central metadata server, Orion scales only to a
small number of clients. Orion omits an evaluation of server fail-over
and recovery (\sys's is in
\S\ref{sec:availability}). Tachyon~\cite{tachyon} aims to circumvent
replication overhead by leveraging the concept of \emph{lineage},
where lost output is recovered by re-executing application code that
created the output. However, to do so, Tachyon requires applications
to use its complex data lineage tracking API.

To maximize NVM utility, we need to design for a scenario where system
call and networking overheads are high compared to storage
access. Hence, \sys eliminates kernel overhead for file operations
and localizes storage operations for both data and meta-data. IO
incurs a single operation to the nearest cache replica in the common
case, without requiring dedicated metadata servers or a distributed
hash to balance load. For scalability, we need to enforce data
consistency locally, which \ccnvm tackles with the help of
leases. \Sys shows how we can realize all the longstanding desirata of
distributed storage at once, without requiring any new APIs.

\paragraph{Leases}~\cite{svm,gray89leases} have long been integral to
providing good performance for distributed file systems by allowing
purely local operations to leased portions of the file name space,
with linearizable semantics.  Read-only leases are a common design
pattern~\cite{afs,sprite,burrowsphd,nfsv4}, but some research systems
have explored using both read and write leases in a similar manner to
Assise.  A prominent example is Berkeley xFS~\cite{xfs}, which
maintained a local block-level update log at each node, written as a
software RAID 5/6 partitioned across other nodes.  Assise differs from
xFS by using an operational log, replicating rather than striping the
log, and by doing update coalescing.

\section{\Sys Design}\label{sec:design}

\begin{figure*}[t]
  \centering
  \begin{subfigure}{.49\textwidth}
    \includegraphics[width=.99\columnwidth]{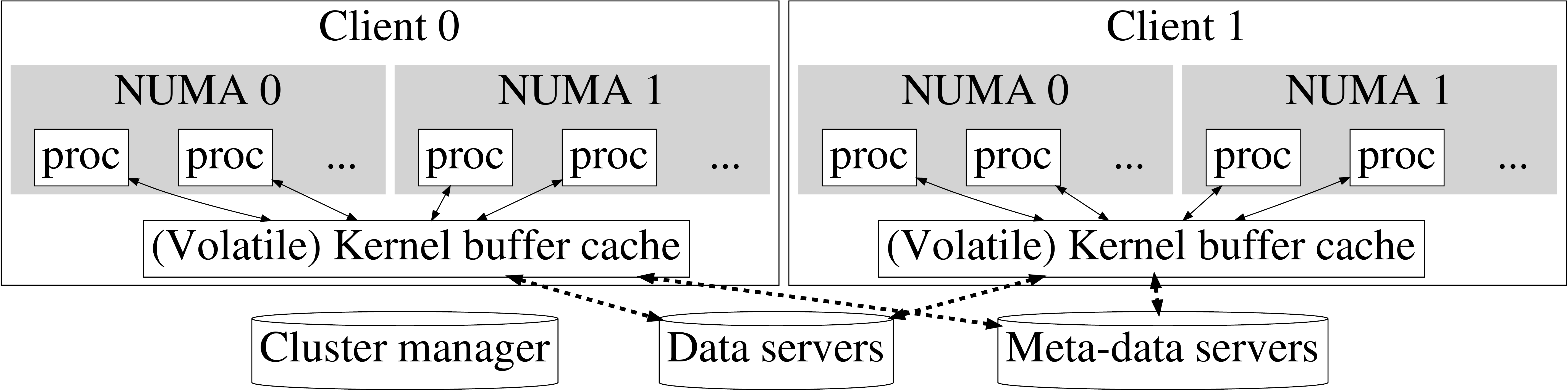}
    \caption{Disaggregated distributed file system (NFS, Ceph, \ldots).
    }
    \label{fig:overview_trad}
  \end{subfigure}
  \begin{subfigure}{.49\textwidth}
    \raggedleft
    \includegraphics[width=.99\columnwidth]{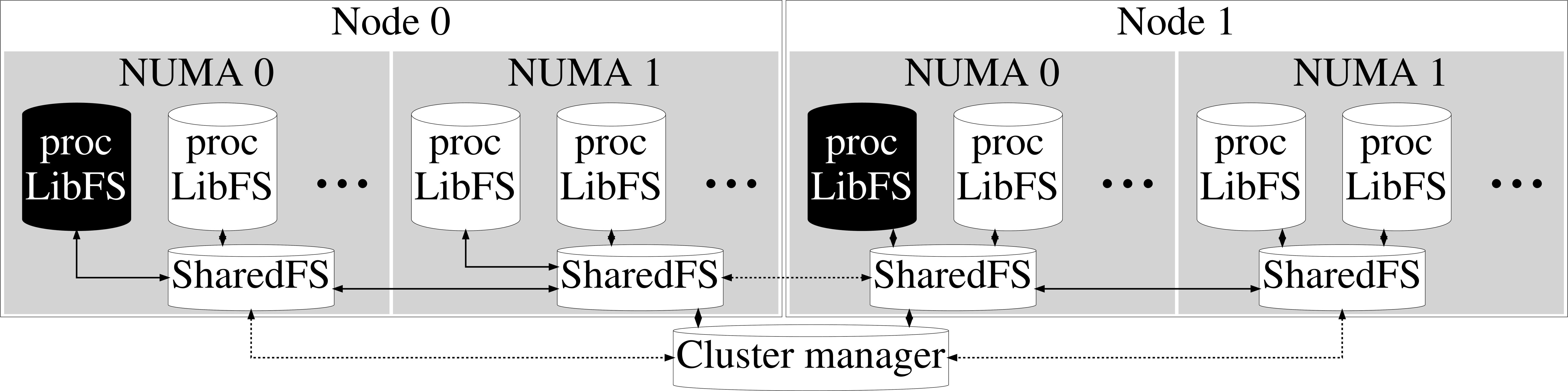}
    \caption{\sys.}
    \label{fig:overview_assise}
  \end{subfigure}
  \caption{Distributed file system state coordination. Arrow = RPC/system
    call. Cylinder = persistence. Black = cache replica.}
  \label{fig:overview}
\end{figure*}

Figure~\ref{fig:overview} contrasts the cache coordination architecture of
disaggregated file systems and \sys. Each subfigure shows two dual-socket nodes
executing a number of application processes sharing a distributed file system. Both
designs use a replicated cluster manager for membership management and failure
detection, but they diverge in all other respects.

\paragraph{Disaggregated file systems} first partition available cluster nodes into
clients and servers. Clients cache file system state in a volatile kernel buffer
cache that is shared by processors across sockets (NVM-NUMA) and accessed via
expensive system calls (NVM-kernel). Persistent file system state is stored in NVM on
remote servers. For persistence and consistency, clients thus have to coordinate
updates with replicated storage and meta-data servers via the network (NVM-RDMA) with
higher latency than local NVM. Data is typically distributed at random over
replicated storage servers for simplicity and load balance~\cite{ceph}. The overhead
of updating a large set of disaggregated storage nodes atomically means that (crash)
consistency is often provided just for meta-data, which is centralized.

\paragraph{\Sys} avoids disaggregated servers and instead uses \ccnvm to coordinate
linearizable state among processes. Processes access cached file system state in
colocated NVM directly via a library file system (\libfs), which may be replicated
for fail-over (2 \libfs cache replicas shown in black). \ccnvm coordinates \libfs{}es
hierarchically via per-socket daemons (\kernfs{}) and the cluster manager.

\paragraph{Crash consistency modes in \sys.} \Sys supports two crash consistency
modes: optimistic or pessimistic~\cite{optfs}. Mount options specify the chosen crash
consistency mode. When pessimistic, \texttt{fsync} forces immediate, synchronous
replication and all writes prior to an \texttt{fsync} persist across failures. When
optimistic, \sys commits all operations in order, but it is free to delay replication
until the application forces it with a \texttt{dsync} call~\cite{optfs}.  Optimistic
mode provides lower latency persistence with higher throughput, but risks data loss
after crashes that cannot recover locally (\S\ref{sec:recovery}). In either mode,
\sys guarantees a crash-consistent file system with prefix
semantics~\cite{salus}---all recoverable writes are in order and no parts of a prefix
of the write history are missing.
%

We now describe cluster coordination and membership management in \sys
(\S\ref{sec:coordination}). We then detail the IO paths (\S\ref{sec:iopaths}) and
show how \ccnvm interacts with them to provide linearizability and crash consistency
with prefix semantics (\S\ref{sec:ccnvm}). Finally, we describe recovery
(\S\ref{sec:recovery}) and reserve replicas (\S\ref{sec:reserve}).

\subsection{Cluster Coordination and Failure Detection}\label{sec:coordination}

Like disaggregated file systems, \sys leverages a replicated cluster manager for
storing the cluster configuration and detecting node failures. \Sys uses
the ZooKeeper~\cite{zookeeper} distributed coordination
service as its cluster manager.

\paragraph{Cluster coordination.} Each \kernfs instance in \sys registers with the
cluster manager. Applications access the file system via a \libfs dynamically linked
into each process. In our prototype, the system administrator decides which \kernfs
replicates which parts of the cached file system namespace at the granularity of a
subtree; the cluster manager records this mapping. When a subtree is first accessed,
\libfs{}es contact their local \kernfs, which consults the cluster configuration and
sets up an RDMA replication chain from \libfs through the subtree's cache
replicas. For each chain, cache replicas allocate a configurable amount of NVM for
replication (\S\ref{app:log}).


\paragraph{Failure detection.} The cluster manager sends heartbeat messages to each
active \kernfs once every second. If no response is received after a timeout, the
node is marked failed and all connected \kernfs are notified.  When the node comes
back online, it contacts the cluster manager and initiates recovery
(\S\ref{sec:recovery}). We leave it as future work to integrate more advanced failure
detectors, e.g., \cite{So:2007:LBF:1272996.1273008,leners11sosp-falcon}.

\subsection{IO Paths}\label{sec:iopaths}

Application IO interacts first with \sys caches. To enable low tail latency IO, \sys
does not rely on a shared kernel buffer cache. Instead, \libfs caches file system
state first in process-local memory; file operations are function calls that
implement the POSIX API. The \libfs cache uses both NVM and DRAM. NVM stores updates,
while DRAM is used to cache read-only state. We now discuss replicated cache
operation upon IO, including eviction and access permissions. Cache coherence is
discussed in \S\ref{sec:ccnvm}.

\vspace*{-.3ex}
\paragraph{Write path.} Writes in \sys occur in two stages: 1. \libfs directly writes
to a process-local cache in NVM. To efficiently support writes of any granularity,
the write cache is an \emph{update log} (\S\ref{sec:ccnvm}), rather than a block
cache. 2. To outlive node failures, the update log is replicated (on \texttt{fsync}
when \emph{pessimistic}, on \texttt{dsync} when \emph{optimistic}) by \libfs to
reserved NVM of the next cache replica along the replication chain via RDMA
(kernel-bypass). The final replica in the chain sends an acknowledgment back along
the chain to indicate that replication completed successfully, and the
\texttt{fsync}/\texttt{dsync} can return.



\vspace{-.3ex}
\paragraph{Read path.}
\libfs first checks the process-local (DRAM and NVM) cache for the requested data. If
not found, it checks the closest cache replica of the corresponding subtree as a
read-shared second level cache. If not found there, \libfs checks reserve replicas
(if any) and, in parallel, cold storage (not shown in
Figure~\ref{fig:overview_assise}). 
%
Reads from remote (NUMA/RDMA) nodes and cold storage are cached in process-local
DRAM. \libfs prefetches up to 256KB from cold storage and up to 4KB from remote
NVM. For small ($<$ 4 KB) remote reads, \libfs first fetches the requested data and
then prefetches the remainder. This minimizes small read latency while improving the
performance of workloads with spatial locality.


\vspace{-.3ex}
\paragraph{Cache eviction.} When a \libfs private cache fills, it replicates any
unreplicated writes and initiates batched eviction on each cache replica along the
replication chain via RDMA RPC. Eviction is done in least-recently-used (LRU)
fashion, first to the \kernfs read-shared caches, then to cold storage.  Each replica
along the chain evicts in parallel and acknowledges when eviction is finished. This
ensures that all replicas cache identical state for fast failover.

Direct stores to NVM on another socket have overheads due to cross-socket hardware
cache coherence, limiting throughput~\cite{swanson_asplos}. Since \ccnvm provides
cache coherence, \sys can bypass hardware cache coherence by using DMA~\cite{ioat}
when evicting to NVM-NUMA. This yields up to 30\% improvement in cross-socket file
system write throughput (\S\ref{sec:scalability}).

\vspace{-.3ex}
\paragraph{Permissions and kernel bypass.} \Sys assumes a single administrative
domain with UNIX file and directory ownership and permissions.
\kernfs enforces that \libfs caches only authorized data, by checking permissions and
data integrity upon eviction and enforcing permissions on reads.
To minimize latency of node-local
\kernfs cache reads, \sys allows read-only mapping of authorized parts of the \kernfs
cache into the \libfs address space. \libfs caches and \kernfs mappings are
invalidated when files or directories are closed and whenever contents are evicted
from the cache.

\subsection{Crash consistent cache coherence with \ccnvm}\label{sec:ccnvm}




\ccnvm provides distributed cache coherence with linearizability when sharing file
system state among processes and with prefix semantics upon a crash.


\paragraph{Crash consistency with prefix semantics.} To provide prefix crash
consistency, \ccnvm tracks write order via the update log in process-local NVM. Each
POSIX call that updates state is recorded, in order, in the update log. When
chain-replicating, \ccnvm leverages the ordering guarantees of (R)DMA to write the
log in order to replicas. This ensures that file system updates are persisted and
replicated atomically and that a prefix of the write history can be recovered
(\S\ref{sec:recovery}).

When in optimistic mode, \sys might coalesce updates to save network bandwidth. To
provide prefix semantics in optimistic mode, \ccnvm wraps each batch of replicated,
coalesced file system operations in a Strata transaction~\cite{strata}. This ensures
that replicated batches are applied atomically, in the event of crashes during
replication.

\paragraph{Sharing with linearizability.} \ccnvm serializes concurrent access to
shared state by untrusted \libfs{}es and recovers the same serialization after a
crash via leases~\cite{gray89leases}. Leases provide a simple, fault-tolerant
mechanism to delegate access. They function similarly to reader-writer locks, but can
be revoked (to allow another process to get a turn) and expire after a timeout (after
which they may be reacquired). In \ccnvm, leases are used to grant shared read or
exclusive write access to a set of files and directories---multiple read leases to
the same set may be concurrently valid, but write leases are exclusive.
Read\-er/\-writ\-er semantics efficiently support shared files and directories that
are read-mostly and widely used, but also write-intensive files and directories that
are not frequently shared. \ccnvm also supports a \emph{subtree lease} that includes
all files and directories at or below a particular directory.  A subtree lease holder
controls access to files and directories within that subtree.  For example, a \libfs
with an exclusive subtree lease on {\tt /tmp/\-bwl-\-ssh/} can recursively create and
modify files and directories within this subtree.

Leases must be acquired by \libfs from \kernfs via a system call before \libfs can
cache the data covered by the lease in process-local memory. \Sys does this upon
first IO; leases are kept until they are revoked by \kernfs.  This occurs when
another \libfs wishes access to a leased file or when a \libfs instance crashes or
the lease times out. Revocation incorporates a grace period in which the current
lease holder can finish its ongoing IO before releasing contended leases. \kernfs
enforces that the private update log and dirty cache entries of the lease holder are
clean and replicated before the lease is transferred. \kernfs logs and replicates
each lease transfer in NVM for crash consistency.  A \libfs may overlap IO with
\kernfs lease replication until \texttt{fsync}/\texttt{dsync}.


\paragraph{Hierarchical coherence.}
To localize coherence enforcement, leases are delegated hierarchically. The cluster
manager is at the root of the delegation tree, with \kernfs{}es as children, and
\libfs{}es as leaves (cf. Figure~\ref{fig:overview_assise}). \libfs{}es request
leases first from their local \kernfs. If the local \kernfs is not the lease holder,
it consults the cluster manager. If there is no current lease holder, the cluster
manager assigns the lease to the requesting \kernfs, which delegates it to the
requesting \libfs and becomes its \emph{lease manager}. If a lease manager already
exists for the requested directory or file, \kernfs forwards the request to the
manager and caches the lease manager's information (leased namespace and expiration
time of lease). The cluster manager expires lease management from \kernfs{}es every 5
seconds. This allows \ccnvm to migrate lease management to the \kernfs that is local
to the \libfs{}es requesting them, while preventing leases from changing managers too
quickly.


The hierarchical structure allows \ccnvm to minimize network communication and thus
lease delegation overhead. Multiple \libfs on the same node or socket require only
local communication with their \kernfs in the common case. This structure maps well
to the data sharding employed by many distributed applications
(\S\ref{sec:scalability}).


\subsection{Fail-over and Recovery}\label{sec:recovery}

\Sys caches file system state with persistence in local NVM, which it can use for
fast recovery. \Sys optimizes recovery performance according to crash prevalence.

\paragraph{\libfs recovery.} An application process crashing is the most common
failure scenario. In this case, the local \kernfs simply evicts the dead \libfs
update log, recovering all completed writes (even in optimistic mode) and then
expires its leases. Log-based eviction is idempotent~\cite{strata}, ensuring
consistency in the face of a system crash during eviction. The crashed process can be
restarted on the local node and immediately re-use all file system state. The \libfs
DRAM read-only cache has to be rebuilt, with minimal performance impact (\S\ref{sec:availability}).

\paragraph{\kernfs recovery.} Another common failure mode is a reboot due to an OS
crash. In this case, we can use NVM to dramatically accelerate OS reboot by storing a
checkpoint of a freshly booted OS. After boot, \sys can initiate recovery for all
previously running \libfs instances, by examining the \kernfs log stored in NVM.


\paragraph{Cache replica fail-over.} 
To avoid waiting for node recovery after a power failure or hardware problem, we
immediately fail-over to a cache replica. The replica's \kernfs takes over lease
management from the failed node, using the replicated \kernfs log to re-grant leases
to any application replicas.  The new instances will see all IO that preceded the
most recently completed \texttt{fsync}/\texttt{dsync}.

Writes to the file system can invalidate cached data of the failed node during its
downtime. To track writes, the cluster manager maintains an epoch number, which it
increments on node failure and recovery. All \kernfs instances are notified of epoch
updates. All \kernfs instances share a per-epoch bitmap in a sparse file indicating
what inodes have been written during each epoch. The bitmaps are deleted at the end
of an epoch when all nodes have recovered.

\paragraph{Node recovery.} When a node crashes, the cluster manager makes sure that
all of the node's leases expire before the node can rejoin. When rejoining, \sys
initiates \kernfs recovery. A recovering \kernfs contacts an online \kernfs to
collect relevant epoch bitmaps. \kernfs then invalidates every block from every file
that has been written since its crash.  This simple protocol could be optimized, for
instance, by tracking what blocks were written, or checksumming regions of the file
to allow a recovering \kernfs to preserve more of its local data.  But the table of
files written during an epoch is small and quickly updated during file system
operation, and our simple policy has been sufficient.

\subsection{Reserve Replicas}\label{sec:reserve}


To fully exploit the memory hierarchy presented in Table~\ref{tab:storage_hierarchy},
remote NVM can be used as a third-level cache, behind local DRAM and local NVM. 
To do so, we introduce \emph{reserve
  replicas}. Like cache replicas, reserve replicas receive all file system updates
via chain-replication, but leverage a different data migration policy. Reserve
replicas track the LRU chain for a specified ``third-level'' portion beyond the
\libfs and \kernfs caches. Reserve replicas evict their third-level, rather than
second-level data to their colocated \kernfs cache.


%
Cache replicas can read from reserve replicas via RDMA with lower latency and higher
bandwidth than cold storage (NVM-RDMA versus SSD in
Table~\ref{tab:storage_hierarchy}). Applications do not run on reserve replicas in
the common case. In the rare case of a failure cascade bringing down all cache
replicas, processes can fail-over to reserve replicas, albeit with reduced short-term
performance. After fail-over, reserve replicas become cache replicas and hot data
must be migrated back into local NVM.


\section{Implementation}\label{sec:impl}



\Sys uses \emph{libpmem}~\cite{pmemio} for persisting data on NVM and
\emph{libibverbs} for RDMA operations in userspace.  \Sys intercepts
POSIX file system calls and invokes the corresponding \libfs
implementation of these functions in
userspace~\cite{syscall-intercept}. The \sys implementation consists
of 28,982 lines of C code (LoC), with \libfs and \kernfs using 16,515
and 6,563 LoC, respectively. The remaining 5,904 LoC contain utility
code, such as hash tables and linked lists. \kernfs runs in its own
user-level process, communicating with \libfs{}es via shared
memory~\cite{barrelfish}. \sys links to Strata for cold storage in SSD
and HDD.

\Sys uses Intel Optane DC PMM in App-Direct mode.  App-Direct makes
NVM visible as a range of persistent memory and is the fastest way to
access NVM, but it requires software support (present in the
prototype). Software-transparent modes have weaker persistence or
performance properties~\cite{optanedc_perf}. Memory mode integrates
NVM as \emph{volatile} memory, using DRAM as a hardware-managed level
4 cache. Sector mode exposes NVM as a disk, with the attendant IO
amplification and disk driver overheads.

\subsection{Efficient Network IO with RDMA}

\Sys makes efficient use of RDMA. For lossless, in-order data transfer
among nodes, \sys uses RDMA reliable connections (RCs). RCs have low
header overhead, improving throughput for small
IO~\cite{rdma_performance, herd}. RCs also provide access to one-sided
verbs, which bypass CPUs on the receiver side, reducing message
transfer times~\cite{farm, pilaf} and memory copies~\cite{tailwind}.


\paragraph{Log replication.} Logs are naturally suited for one-sided
RDMA operations and \sys uses RDMA writes for log
replication. Replication operations typically require only one write,
reducing header and DMA overheads~\cite{rdma_performance}. The only
exceptions are when the remote log wraps around or when the local log
is fragmented (due to coalescing), such that it exceeds the hardware
limit for scatter-gather writes.

\paragraph{Persistent RDMA writes.} The RDMA specification does not
define the persistence properties of remote NVM access via RDMA. In
current practice, the remote CPU is required to flush any written data
from the remote processor cache to NVM. \Sys flushes all writes via
the CLWB and SFENCE instructions on each replica, before acknowledging
successful replication. In the future, it is likely that enhancements
to PCIe will allow RDMA NICs to bypass the processor cache and write
directly to NVM to provide persistence~\cite{hyperloop}.

\paragraph{Remote NVM reads.}
\Sys reads remote data via RPC. To keep the request sizes small, \sys
identifies files using their inode numbers instead of their
path. 
As an optimization, DRAM read cache locations are pre-registered with
the NIC. This allows the remote node to reply to a read RPC by RDMA
writing the data directly to the requester's cache, obviating the need
for an additional data copy.
\section{Evaluation}\label{sec:eval}


We evaluate \sys's common-case as well as its recovery performance,
and break down the performance benefits attained by each system
component. We compare \sys to three state-of-the-art distributed file
systems that support NVM and RDMA. Our experiments rely on several
microbenchmarks and Filebench~\cite{filebench} profiles, in addition
to several real applications, such as LevelDB, Postfix, and
MinuteSort. Our evaluation answers the following questions:


\begin{compactitem}[\labelitemi]
\item \textbf{IO latency and throughput breakdown
    (\S\ref{sec:microbench}).} What is the hardware IO performance of
  a storage hierarchy with colocated NVM
  (Table~\ref{tab:storage_hierarchy})?  How close to this performance
  do the file systems operate under various IO patterns? What are the
  sources of overhead?

\item \textbf{Cloud application performance (\S\ref{sec:appbench}).}
  What is the performance of cloud applications with various
  consistency, latency, throughput, and scalability requirements? By
  how much can a reserve replica improve read latency? By how much can
  optimistic crash consistency improve write throughput by eliminating
  redundant writes? What is the overhead of \sys's POSIX API
  implementation versus hand-tuned direct use of colocated NVM?


\item \textbf{Availability (\S\ref{sec:availability}).} How quickly
  can applications recover from various failure scenarios?

\item \textbf{Scalability (\S\ref{sec:scalability}).} How do multiple
  processes sharing the file system scale? By how much can \sys's
  hierarchical coherence improve multi-node and multi-socket
  scalability?






\end{compactitem}


\paragraph{Testbed.} Our experimental testbed consists of 5$\times$
dual-socket Intel Cascade Lake-SP servers running at 2.2GHz, with a
total of 48 cores (96 hyperthreads), 384 GB DDR4-2666 DRAM, 6 TB Intel
Optane DC PMM, 375 GB Intel Optane DC P4800X series NVMe-SSD, and a 40
GbE ConnectX-3 Mellanox Infiniband NIC. To leverage all 6 memory
channels per processor, there are 6 DIMMs of DRAM and NVM per socket.
NVM is used in App-Direct mode (\S\ref{sec:impl}).
All nodes run Fedora 27 with Linux kernel version 4.18.19 and are
connected via an Infiniband switch.

\paragraph{Hardware performance.} We first measure the achievable IO
latency and throughput for each memory layer in our testbed server. We
do this by using sequential IO and as many cores of a single socket as
necessary. We measure DRAM and NVM (App-Direct) latency and throughput
using Intel's memory latency checker~\cite{intel_mlc}. NVM-RDMA
performance is measured using RDMA read and write-with-immediate (to
flush remote processor caches) operations to remote NVM. SSD
performance is measured using \verb+/dev/nvme+ device files. The IO
sizes that yielded maximum performance are 64 B for DRAM, 256 B for
NVM, and 4 KB for SSD.  Table~\ref{tab:storage_hierarchy} shows these
results. The measured IO performance for DRAM, NVM, and SSD match the
hardware specifications of the corresponding devices
and is confirmed by others~\cite{optanedc_perf}.
NVM-RDMA throughput matches the line rate of the NIC. NVM-RDMA write
latency has to invoke the remote CPU (to flush caches) and is thus
larger than read latency. We now investigate how close to these limits
each file system can operate.



\begin{table}
  \centering
  \footnotesize
  \begin{tabular}{p{3.9cm}rrrr}
    Feature & Assise & Ceph & NFS & Octopus \\
    \hline
    Cache replication & \checkmark & & & \\
    Local consistency & \checkmark & & & \\
    Linearizability & \checkmark & & & \\
    Data crash consistency & \checkmark & & & \\
    Kernel-bypass & \checkmark & & & \checkmark \\
    Replication & \checkmark & \checkmark & & \\
    RDMA & \checkmark & \checkmark & \checkmark & \checkmark
  \end{tabular}
  \caption{Features of evaluated distributed file systems.\vspace{-2ex}}%
  \label{tab:fs_features}
\end{table}

\paragraph{State-of-the-art.} Table~\ref{tab:fs_features} shows
performance-relevant features of the state-of-the-art and \sys. We can
see that no open-source distributed file system provides all of \sys's
features. Hence, a direct performance comparison is difficult.  We
perform comparisons against the Linux kernel-provided NFS version
4~\cite{nfsv4} and Ceph version 14.2.1~\cite{ceph} with
BlueStore~\cite{bluestore}, both retrofitted for RDMA, as well as
Octopus~\cite{octopus}. We cannot directly compare with
Orion~\cite{orion} as it is not publicly available, but we emulate its
behavior where possible. Only Ceph provides availability via
replicated object storage daemons (OSDs), delegating meta-data
management to a (potentially sharded) meta-data server (MDS). 
Octopus and NFS do not support replication for availability and thus
gain an unfair performance advantage over \sys. However, \sys beats
them even while replicating for availability, showing that both
features can be had when leveraging colocated NVM.

No other file system supports persistent caches and
their consistency semantics are often weaker than \sys's. \Sys provides data
crash consistency, while both Ceph/Bluestore and Octopus provide only
meta-data crash consistency~\cite{consistency_wo_ordering}.
For NFS, crash consistency is determined by the underlying file
system. We use EXT4-DAX~\cite{ext4dax}, which also provides only
meta-data crash consistency.
When sharing data, NFS provides \emph{close-to-open
  consistency}~\cite{nfsv4}, while Octopus and Ceph provide ``stronger
consistency than NFS''~\cite{cephfs-posix}, and \sys provides
linearizability which is stronger than Octopus and Ceph's guarantee.
In all performance comparisons, \sys provides stronger consistency
than alternatives.  Ceph is the closest comparison point.

\paragraph{File system compliance tests.} We tested \sys using
xfstests~\cite{xfstests} and CrashMonkey~\cite{crashmonkey}. \Sys
passed all 75 generic xfstests that are recommended for
NFS~\cite{nfstests}. NFSv4.2 and Ceph v14.2.1 pass only 71 and 69 of
these tests, respectively. In part, this is due to their weaker
consistency model (\S\ref{app:tests}). \Sys also successfully passes
CrashMonkey tests, runs all existing Filebench profiles, passes all
unit tests for the LevelDB key-value store, and passes MinuteSort
validation.

\subsection{Testbed configuration}

\paragraph{Machines.} Each experiment specifies the number ($\geq 2$) of
testbed machines used. By default, machines are used as cache replicas
in \sys, as a pool of storage nodes in Octopus, and as OSD and MDS
replicas in Ceph. NFS uses only one machine as server, the rest as
clients. We place applications on cache replicas for \sys, on OSD
replicas for Ceph, on storage nodes for Octopus, and on clients for
NFS. \Sys's and Ceph's cluster managers run on 2 additional testbed
machines (NFS and Octopus do not have cluster managers).
%
%

\paragraph{Network.} We use RDMA for the NFS client-server connection.
Ceph provides its client-side file system via the Ceph kernel driver
and uses IP over Infiniband, which was the fastest configuration (we
also tried FUSE and Accelio~\cite{accelio}). \Sys and Octopus use RDMA
with kernel-bypass.

\paragraph{Storage and caches.} For maximum
efficiency, all file systems use NVM in App-Direct mode to provide
persistence (cylinders in Figure~\ref{fig:overview}) and DRAM when
persistence is not needed (e.g., kernel buffer cache). We investigate
Ceph and NFS performance using NVM in memory mode for volatile caches
and found it to degrade throughput by up to 25\% versus DRAM. For
efficient access to NVM, Ceph OSDs use Bluestore and NFS servers use
EXT4-DAX. Octopus uses FUSE to provide its file system interface to
applications in \emph{direct IO} mode to NVM, bypassing the kernel
buffer cache~\cite{octopus_repo}.

To evaluate a breadth of cache behaviors with limited
application data set sizes, we limit the fastest cache size for all
file systems to 3GB. For Ceph and NFS, we limit the kernel buffer
cache to 3GB. For \sys, we partition the \libfs cache into a 1GB NVM
update log and a 2GB DRAM read cache (the \kernfs second-level cache may
use all NVM available), and we run \sys in pessimistic mode. The
impact of log sizing is excluded from this evaluation but detailed in
\S\ref{app:log}.

\subsection{Microbenchmarks}\label{sec:microbench}

\begin{figure}[t]
  \centering
  \begin{subfigure}{\columnwidth}
    \includegraphics[width=0.9\columnwidth]{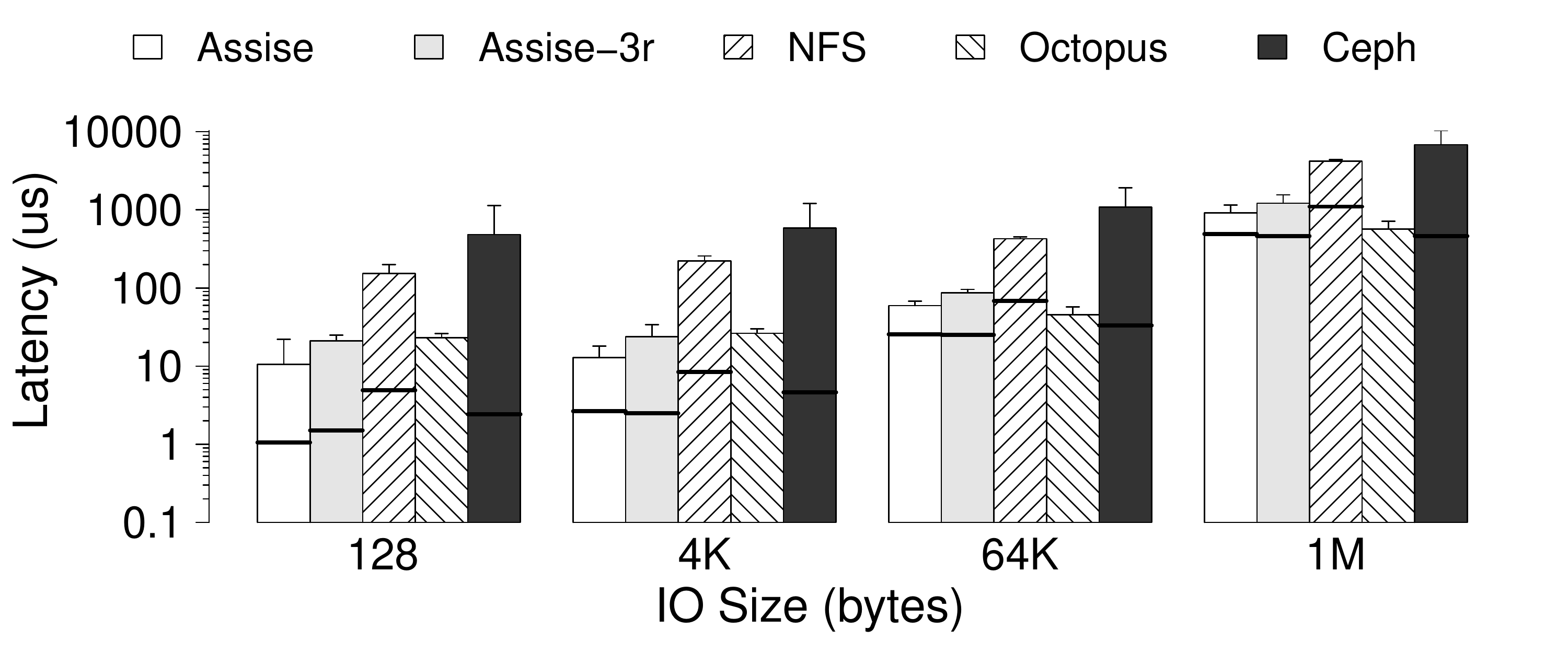}
    \subcaption{Sequential write. {\tt Write} latency is solid line, {\tt fsync} is bar height.
	}
    \label{fig:micro_write_latency}
  \end{subfigure}
  
  \begin{subfigure}{\columnwidth}
    \includegraphics[width=0.9\columnwidth]{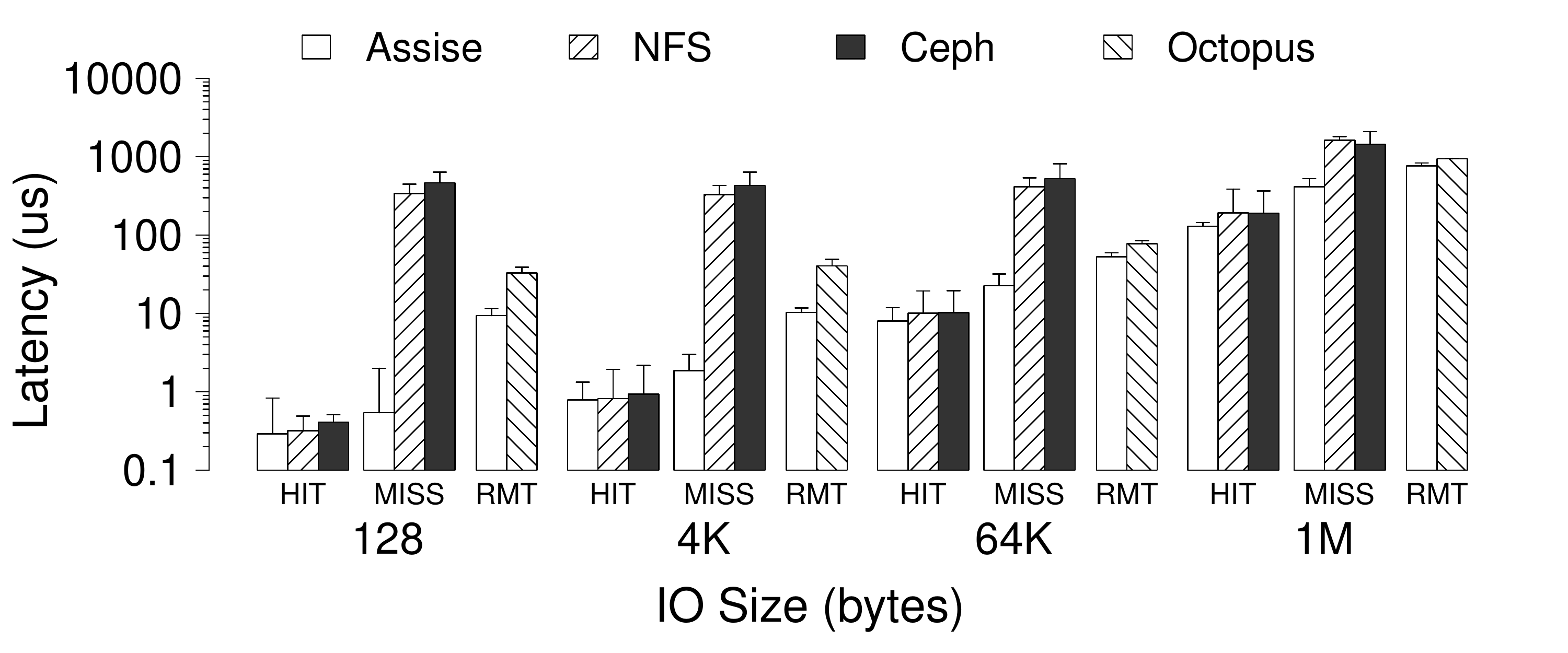}
    \subcaption{Read latencies for cache hits, misses, and remote (RMT) misses.}
    \label{fig:micro_read_latency}
  \end{subfigure}

  \caption{Average and 99\%ile (error bar) IO latencies. Log scale.}
  

\end{figure}

\paragraph{Average and tail write latency.} We compare synchronous
write latencies on an otherwise idle cluster with 2 machines (except
\sys-3r which uses 3 machines). Synchronous writes involve
\texttt{write} calls that operate locally (except for Octopus where
writes may be remote), and \texttt{fsync} calls that involve remote
nodes for replication (\sys, Ceph) and/or persistence (Ceph,
NFS). Each experiment appends 1 GB of data into a single file, and we
report per-operation latency. In this case, the file size is smaller
than each file system's cache size, so no evictions occur---with
gigabytes of cache capacity, this is common for latency-sensitive
write bursts.

Figure~\ref{fig:micro_write_latency} shows the average and 99th
percentile sequential write latencies over various common IO sizes
(random write latencies are similar for all file systems).
%
%
We break writes down into \texttt{write} (solid line) and
\texttt{fsync} call latencies (bar). Octopus' \texttt{fsync} is a
no-op. \Sys's local write latencies match that of
Strata~\cite{strata}.  \Sys's average write latency for 128B two-node
replicated writes is only 8\%
higher than the aggregate latencies of the required local and NVM-RDMA
writes (cf.  Table~\ref{tab:storage_hierarchy}). Three replicas
(\sys-3r) increase \sys's overhead to 2.2$\times$ due to
chain-replication with sequential RPCs. The 99th percentile replicated
write latency is up to 2.1$\times$ higher than the average for 2
replicas. This is due to Optane PMM write
tail-latencies~\cite{optanedc_perf}. The tail difference diminishes to
19\% for 3 replicas due to the higher average.

Ceph and NFS use the kernel buffer cache and interact at 4KB block
granularity with servers. 
For small writes, the incurred network IO amplification during
\texttt{fsync} is the main reason for up to an order of magnitude
higher aggregate write latency than \sys. In this case, their
\texttt{write} latency is up to 3.2$\times$ higher than \sys due to
kernel crossings and copy overheads. For large writes ($\geq$ 64 KB), network IO
amplification diminishes but the memory copy required to maintain a
kernel buffer cache becomes a major overhead. Large \texttt{write} latency
is higher than \sys's replicated write latency (and up to
2.7$\times$ higher than \sys's non-replicated \texttt{write} latency), while
synchronous write latency is up to 7.2$\times$ higher than \sys. Ceph
has higher \texttt{fsync} latency than NFS due to replication.

Octopus eliminates the kernel buffer cache and block orientation,
which improves its performance drastically versus NFS and
Ceph. However, Octopus still treats all NVM as disaggregated and uses
FUSE for file IO. Octopus exhibits up to 2.1$\times$ higher latency
than \sys for small (< 64 KB) writes. This overhead stems from FUSE
(around 10$\mu$s~\cite{fuse_perf}) and writing to remote NVM via the
network.
Large writes ($\geq$ 64 KB) amortize Octopus' write overheads and \sys
has up to 1.7$\times$ higher write latency due to \sys's replication
(for availability).  Octopus does not replicate.



%

\paragraph{Average and tail read latency.} Read latency is affected by
whether a read hits or misses in the cache. 
We show both cases by reading a 1GB file with various IO sizes, once
with a warm cache and once with a cold cache. The results are shown in
Figure~\ref{fig:micro_read_latency}. \Sys has a second-layer cache in
\kernfs before going remote, and we report three cases for
\sys. Reads in Octopus are always remote.

We first compare cache-hit latencies (HIT), where \sys is up to 40\%
faster than NFS and 50\% faster than Ceph. \Sys serves data from the
\libfs read cache, while NFS and Ceph use the kernel buffer cache.
%
If \sys misses in the \libfs cache, data may be served from the local
\kernfs (MISS). \sys-MISS incurs up to 3.2$\times$ higher latency than
\sys-HIT due to reading the extent tree index, especially for larger IO
sizes that read a greater number of extents. If \sys misses in both caches,
it has to read from a remote replica (RMT). Due to the use of RDMA
from user-space, \sys-RMT latency is close to that of an RDMA RPC.
When NFS and Ceph miss in the cache, their clients have to fetch from
disaggregated servers, which incurs up to orders of magnitude higher
average and tail latencies than \sys-RMT and \sys-MISS. Ceph performs
worse than NFS due to a more complex OSD read path.

The elimination of a cache hurts Octopus' read performance, because it
has to fetch metadata and data (serially) from remote NVM
(RMT). Octopus' read latency is up to two orders of magnitude higher
than the other file systems hitting in the cache, but up to an order
of magnitude lower than NFS and Ceph missing in the cache. Octopus
does not handle small ($\leq$ 4KB) reads well due to FUSE overhead,
with up to 3.54$\times$ \sys-RMT read latency. This overhead amortizes
for larger reads ($\geq$ 64KB), where Octopus incurs up to
1.46$\times$ the read latency of \sys-RMT. By configuring FUSE to use
the kernel buffer cache for Octopus, we reduce Octopus' read hit
latency to 1.8$\times$ that of \sys-HIT, with the remaining overhead
due to FUSE. However, using the kernel buffer cache inflates write
latencies for Octopus by up to an order of magnitude due to additional
buffer cache memory copies.


\begin{figure}[t]
  \begin{subfigure}[t]{.5\columnwidth}
    \includegraphics[width=\columnwidth]{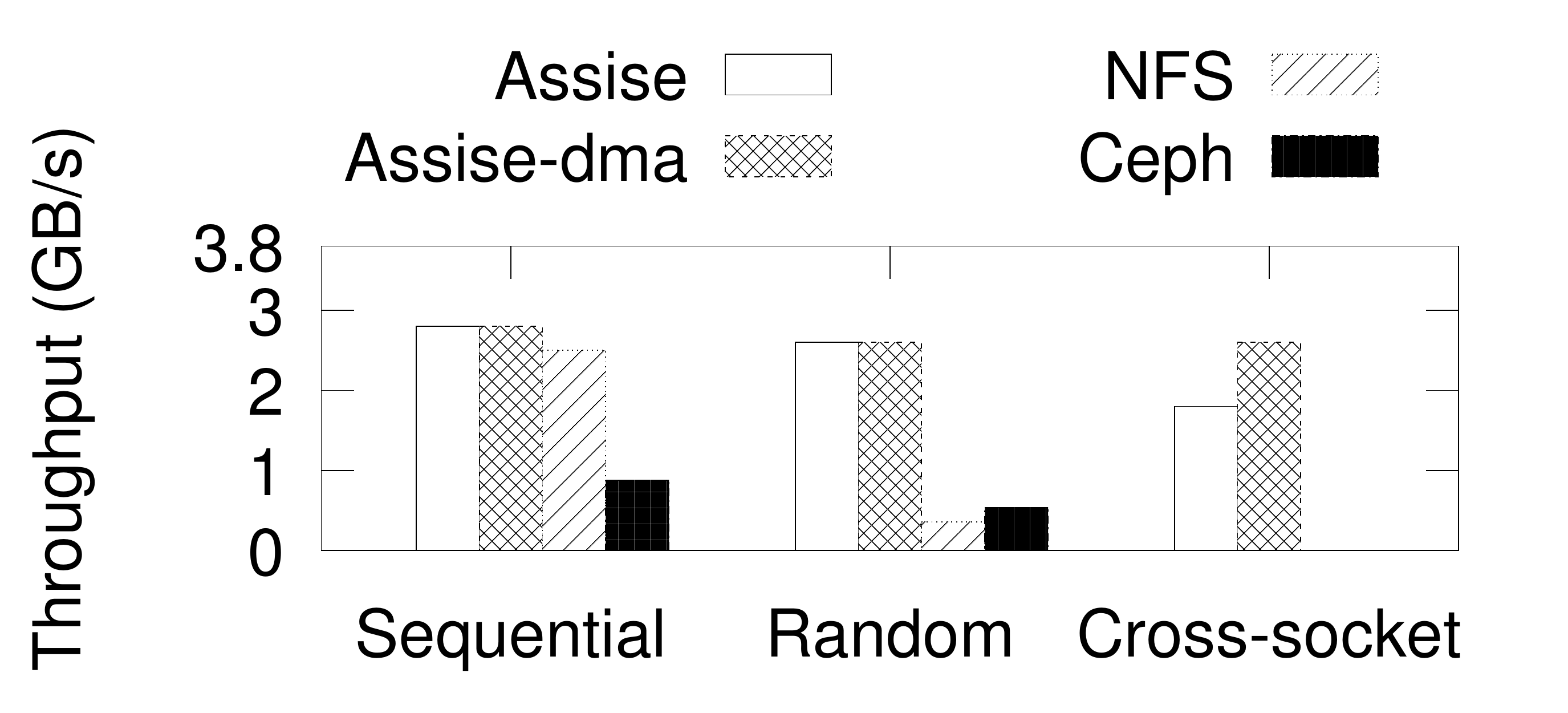}
    \subcaption{Write. 3.8GB/s is NVM-RDMA b/w.
    }
    \label{fig:micro_write_tput}
  \end{subfigure}%
  \begin{subfigure}[t]{.5\columnwidth}
    \includegraphics[width=\columnwidth]{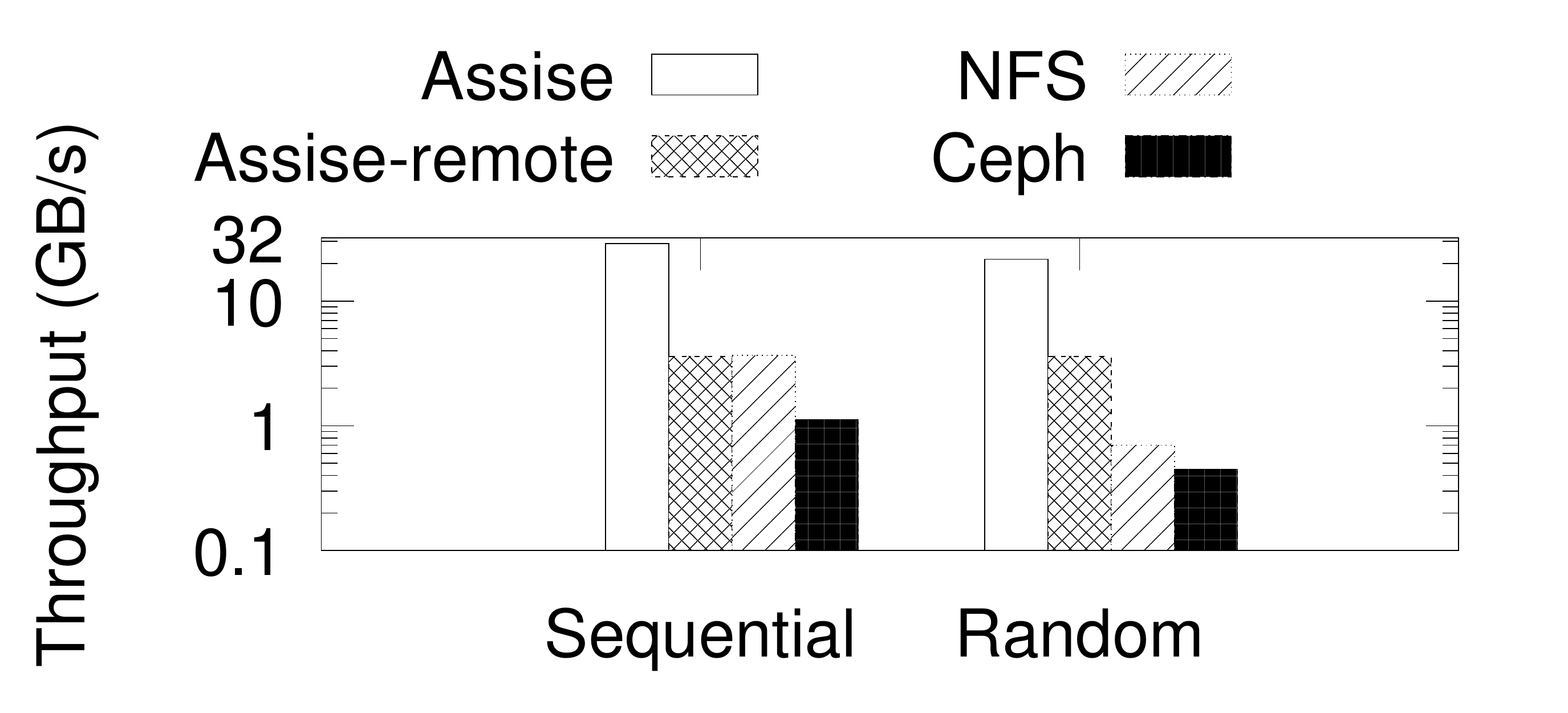}
    \subcaption{Read. 32GB/s is NVM read b/w.
    }
    \label{fig:micro_read_tput}
  \end{subfigure}
  \caption{Average throughput with 24 threads at 4 KB IO size.
	  }
  \label{fig:micro_tput}
\end{figure}

\paragraph{Peak throughput.} Figure~\ref{fig:micro_tput} shows average
throughput of sequential and random IO to a 120 GB dataset (on the
local socket) with 4KB IO size from 24 threads (all cores of one
socket). To evaluate a standard replication factor of 3, we use 3
machines for \sys and Ceph. The dataset is sharded over 24 files, or
one 5 GB file per thread.  \texttt{write} calls are not followed by
\texttt{fsync}
and 
the total amount of accessed data is larger than the cache size,
causing cache eviction on write and cache misses on read. For \sys,
we show cache miss performance from a local and remote \kernfs.
Octopus crashes during this experiment and is not shown.

For sequential writes, \sys and NFS achieve 74\% and 66\% of the
NVM-RDMA bandwidth (cf. Table~\ref{tab:storage_hierarchy}),
respectively, due to protocol overhead for NFS and log header overhead
for \sys. Chain-replication in \sys affects throughput only
marginally.
Ceph replicates in parallel to 2 remote replicas, consuming 3$\times$
the network bandwidth. This reduces its throughput to 31\% of \sys and
35\% of NFS. \Sys achieves similar performance for sequential and
random writes, as \sys's writes are log-structured. NFS and Ceph
perform poorly for random writes due to cache block mis-prefetching
incurring additional reads from remote servers, causing \sys to
achieve 4.8$\times$ Ceph's throughput. NFS throughput is at only 67\%
that of Ceph, which is due to kernel locking overhead.

To quantify the benefit of bypassing hardware cache coherence for
cross-socket writes with DMA, we repeat the benchmark, placing the
target directory on the remote socket. We can see that \sys-dma
attains 44\% higher cross-socket throughput than non-temporal
processor writes (\sys). Sequential and random writes provide
comparable performance. NVM-NUMA writes occur during digestion from
the \libfs update log (local socket) to the shared area (remote
socket). When writing to the local socket, \sys-dma attains identical
throughput to \sys, regardless of pattern.

For local sequential and random reads from the colocated \kernfs
cache, \sys achieves 90\% and 68\%, respectively, of local, sequential
NVM bandwidth. The 10\% difference for sequential reads to local NVM
bandwidth is due to meta-data lookups, while random reads additionally
suffer PMM buffer misses~\cite{optanedc_perf}. \Sys remote reads
attain full NVM-RDMA bandwidth (3.8 GB/s), regardless of access
pattern.
NFS and Ceph are limited by NVM-RDMA bandwidth for all reads and again
have worse random read performance due to prefetching.

\subsection{Application Benchmarks}\label{sec:appbench}

We evaluate the performance of a number of common cloud applications,
such as the LevelDB key-value store~\cite{leveldb}, the Fileserver and
Varmail profiles of the Filebench~\cite{filebench} benchmarking suite,
emulating file and mail servers, and the MinuteSort benchmark. We use
3 machines for LevelDB and Filebench and 5 machines for MinuteSort.

\begin{figure}
  \centering
  \includegraphics[width=.98\columnwidth]{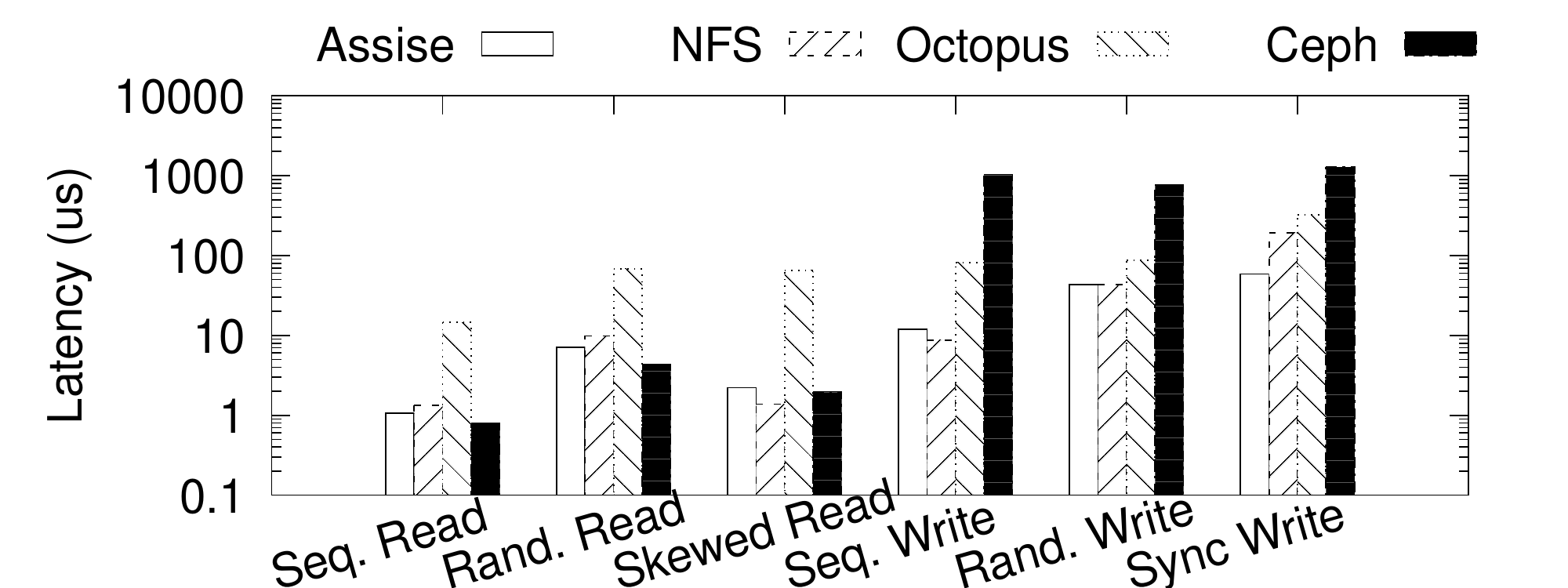}
  \caption{Average LevelDB benchmark latencies. Log scale.}
  \label{fig:leveldb}
\end{figure}

\paragraph{LevelDB.} We run a number of single-threaded LevelDB
latency benchmarks, including sequential and random IO, skewed random
reads with 1\% of highly accessed keys, and sequential synchronous
writes (\texttt{fsync} after each write). All benchmarks use a key
size of 16 B and a value size of 1 KB with a working set of 1M KV
pairs. Figure~\ref{fig:leveldb} presents the average measured
operation latency, as reported by the benchmark.

\Sys, Ceph, and NFS perform similarly for reads, where caching allows
them to operate close to hardware speeds. For non-synchronous writes,
NFS is up to 26\% faster than \sys, as these go to its client kernel
buffer cache in large batches (LevelDB has its own write buffer),
while \sys is 69\% faster than NFS for synchronous writes that cannot
be buffered. Random IO and synchronous writes incur increasing LevelDB
indexing overhead for all results. Ceph performs worse than NFS for
writes because it replicates (as does \sys) and \sys performs
22$\times$ better. Octopus bypasses the cache and thus performs worst
for reads and better only than Ceph for writes, as it does not
replicate.

\begin{figure}
  \centering
  \includegraphics[width=\columnwidth]{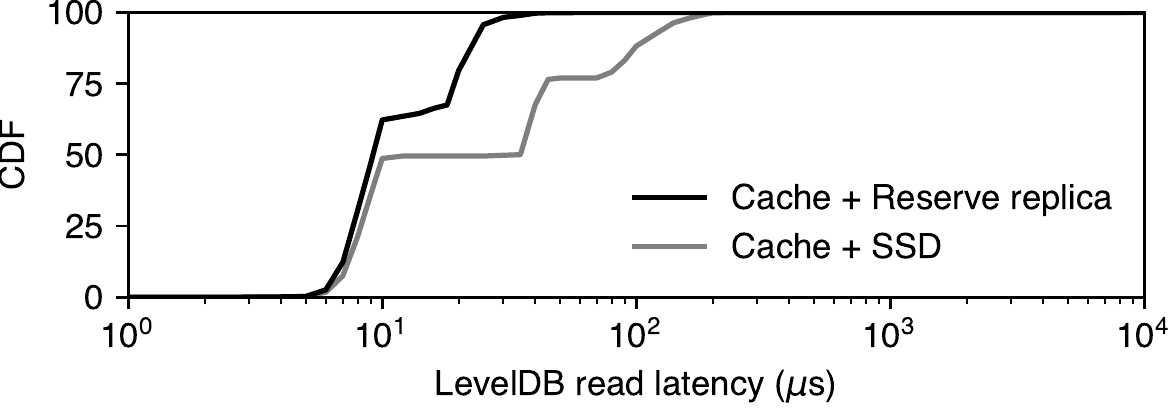}
  \caption{LevelDB random read latencies with SSD/reserve replica.}
  \label{fig:reserve}
\end{figure}

\paragraph{Reserve replica read latency.} Reserve replicas reduce read
latency for cold data by allowing these reads to be served from remote
NVM, rather than cold storage. For this benchmark, we configure \sys
to limit the aggregate (\libfs and \kernfs) cache to 2GB and use the
local SSD for cold storage. We then run the LevelDB random read
experiment with a 3GB dataset. We repeat the experiment 1) with 3
cache replicas and 2) with 2 cache and 1 reserve
replica. Figure~\ref{fig:reserve} shows a CDF of read latencies. The
benchmark accesses data uniformly at random, causing 33\% of the reads
to be cold. Consequently, at the 50th percentile, read latencies are
similar for both configurations (served from cache). At the 66th
percentile, reads in the first setup are served from SSD and have
2.2$\times$ higher latency than reserve replica reads in the second
setup. At the 90th percentile, the latency gap extends to 6$\times$.

\begin{figure}[t]
    \centering
    \includegraphics[width=\columnwidth]{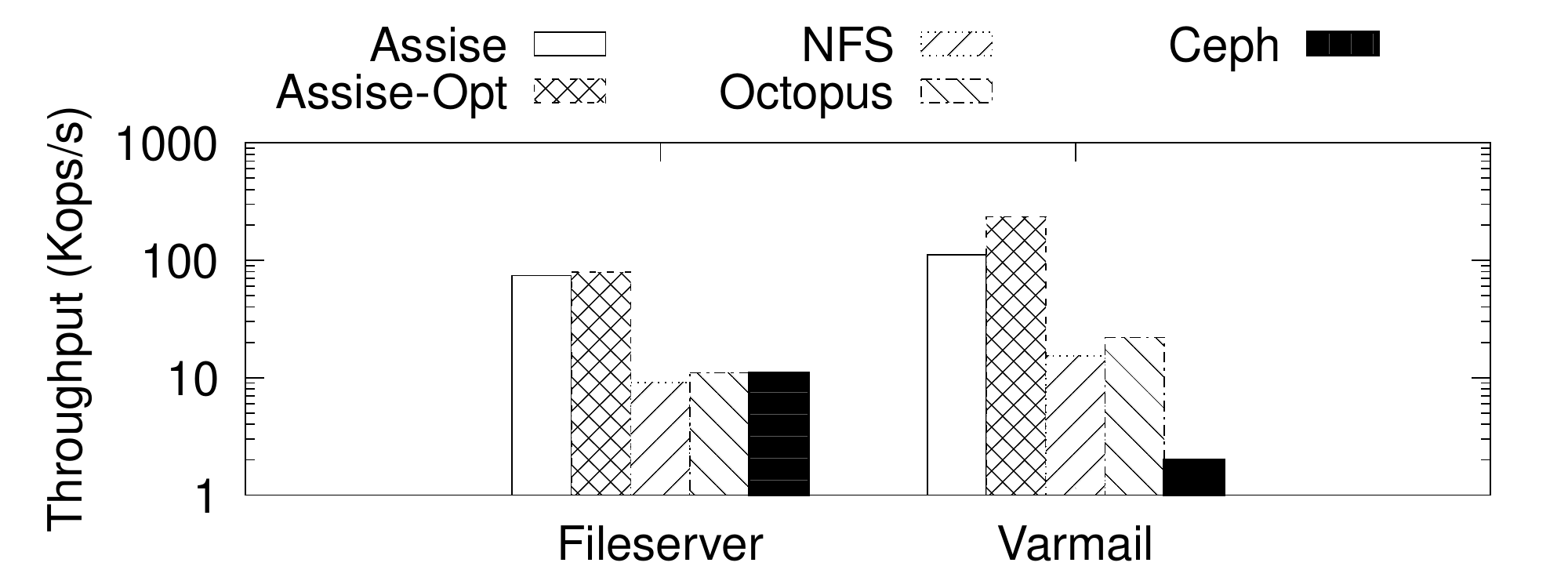}
    \caption{Average Varmail and Fileserver
      throughput. Log scale.
    }
    \label{fig:filebench}
\end{figure}

\paragraph{Filebench.} Varmail and Fileserver operate on a working set
of 10,000 files of 16 KB and 128 KB average size, respectively. Files
grow via 16 KB appends in both benchmarks (mail delivery in
Varmail). Varmail reads entire files (mailbox reads) and Fileserver
copies files. Varmail and Fileserver have write to read ratios of 1:1
and 2:1, respectively. Varmail leverages a write-ahead log with strict
persistence semantics (\texttt{fsync} after log and mailbox writes),
while Filebench consistency is relaxed (no \texttt{fsync}).
Figure~\ref{fig:filebench} shows average measured throughput of both
benchmarks.  \Sys outperforms Octopus (the best alternative) by
6.7$\times$ for Fileserver and 5.1$\times$ for Varmail,
respectively. Ceph performs worse than NFS for Varmail due to stricter
persistence requiring it to replicate frequently and due to MDS
contention, as Varmail is meta-data intensive. 


\paragraph{Optimistic crash consistency.} We repeat this benchmark for
\sys in optimistic mode (\sys{}-Opt) and change Varmail to use
synchronous writes for the mailbox, but non-synchronous writes for the log.
Prefix semantics allow \sys to buffer and eliminate
(\emph{coalesce}~\cite{strata}) the temporary log write without losing
consistency. \sys{}-Opt achieves 2.1$\times$ higher throughput than
\sys. Fileserver has few redundant writes and \sys{}-Opt is only 7\%
faster.

\begin{table}
  \footnotesize
  \centering
  \begin{tabular}{lrrrrrr}
    File system & Processes & Partition [s] & Sort [s] & \textbf{Total [s]} & \textbf{GB/s} \\
    \hline
    \Sys & 160 & 20.3 & 43.0 & \textbf{63.3} & \textbf{5.1} \\
     & 320 & 52.1 & 43.0 & \textbf{95.1} & \textbf{6.7} \\
    NFS & 160 & 60.9 & 79.3 & \textbf{140.2} & \textbf{2.3} \\
     & 320 & 104.1 & 84.2 & \textbf{188.3} & \textbf{3.4} \\
    DAX & 320 & -- & 44.1 & -- & -- \\
  \end{tabular}
  \caption{Average Tencent Sort duration breakdown.\vspace{-1ex}}%
  \label{tab:minutesort}
\end{table}

\paragraph{MinuteSort.} We implement and evaluate Tencent
Sort~\cite{tencentsort}, the current winner of the MinuteSort external
sorting competition~\cite{minutesort}. Tencent Sort sorts a
partitioned input dataset, stored on a number of cluster nodes, to a
partitioned output dataset on the same nodes. It conducts a
distributed sort consisting of 1) a range partition and 2) a mergesort
(cf. MapReduce~\cite{mapreduce}). Step 1 presorts unsorted input files
into ranges, stored in partitioned temporary files on destination
machines. Step 2 reads these files, sorts their contents, and writes
the output partitions. Each step uses one process per partition; the
number of partitions determines the sort parallelism. A distributed
file system stores the input, output, and temporary files, implicitly
taking care of all network operations.

We benchmark the MinuteSort Indy category. Indy requires sorting a
synthetic dataset of 100 B records with 10 B keys, distributed
uniformly at random. Creating a 2GB input partition per process, we
run 160 and 320 processes in parallel, uniformly distributed over 4
machines. MinuteSort does not require replication, so we turn it
off. It calls \texttt{fsync} only once for each output partition,
after the partition is written. We compare a version running a single
\sys file system with one leveraging per-machine NFS mounts. For \sys,
we configure the temporary and output directories to be colocated with
the mergesort processes. We do the same for NFS, by exporting
corresponding directories from each mergesort node. We conduct three
runs of each configuration and report the average. We use the official
competition tools~\cite{minutesort} to generate and verify the input
and output datasets. Table~\ref{tab:minutesort} shows that \sys sorts
up to 2.2$\times$ faster than NFS. Running twice the number of
processes only marginally improves performance, as \sys is
bottlenecked by network bandwidth. If the performance of \sys scales
to the cluster size of the original Tencent Sort (512 machines), we
can estimate that \sys sorts 1.5$\times$ more data per minute than the
world record holder.

To show that \sys's POSIX implementation does not reduce performance,
we modify the sort step to map all files into memory using EXT4-DAX
and use processor loads and non-temporal stores to sort directly in
NVM, rather than using file IO. We can see that the sort phase is 3\%
slower with DAX. libc buffers IO in DRAM to write 4KB at a time to
NVM, performing better than direct, interleaved appends of 100B
records.

\subsection{Availability}\label{sec:availability}

\begin{figure}[t]
  \centering
  \includegraphics[width=\columnwidth]{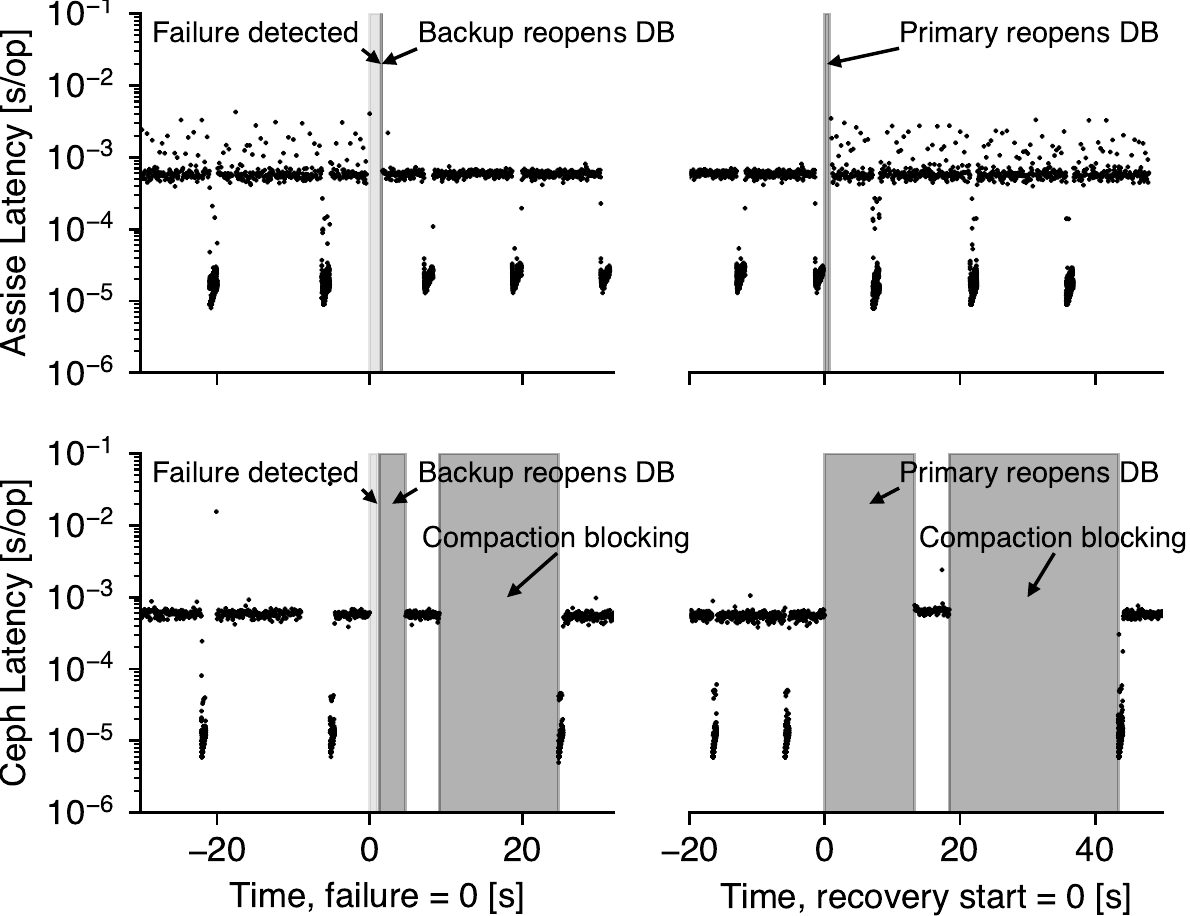}
  \caption{LevelDB operation latency time series during fail-over and
    recovery. Log scale.\vspace{-1ex}}
  \label{fig:failover}
\end{figure}

Ceph and \sys are fault tolerant. We evaluate how quickly these file
systems return an application back to full performance after various
fail-over and recovery situations. To do so, we run LevelDB on the
same dataset (\S\ref{sec:appbench}) with a 1:1 read-write ratio and
measure operation latency before, during, and after fail-over and
recovery. These experiments use 2 machines (primary and
backup). LevelDB initially runs on the primary, where we inject
failures. Failures are detected using a 1 s heartbeat timeout via each
file system's cluster manager. Once a failure is detected, LevelDB
immediately restarts on the backup. We report average results over 5
benchmark runs.

\paragraph{Fail-over to hot backup.} A time series of measured LevelDB
operation latencies during one experiment run is shown in
Figure~\ref{fig:failover}. Pre-failure, we see bursts of low latency
in between stretches of higher latency. This is LevelDB's
steady-state. Bursts show LevelDB writes to its own DRAM log. These
are periodically merged with files when the DRAM log is full, causing
writes that are higher latency (and sometimes blocking with Ceph), as
the writes wait on the log to become available.

We inject a primary failure by killing the primary's file system
daemon (\kernfs for \sys and OSD for Ceph) and LevelDB. During primary
failure, no operations are executed. It takes 1 s to detect the
failure and restart LevelDB on the backup (light shaded box). Due to
unclean shutdown, LevelDB first checks its dataset for integrity
before executing further operations (dark shaded box). For failover,
\sys need only evict the per-process log (up to 1GB) on the backup
cache replica, making fail-over near-instantaneous. LevelDB returns to
full performance in both latency and throughput 230 ms after failure
detection. On Ceph, it takes 3.7 s after failure detection until
further operations are executed. However, LevelDB stalls soon
thereafter upon compaction (further dark shaded box), which involves
access to further files, resulting in an additional 15.6 s delay,
before reaching steady-state. Ceph's long aggregate fail-over time of
23.7 s is due to Ceph losing its DRAM cache, which it rebuilds during
LevelDB restart. \Sys reaches full performance after failure detection
103$\times$ faster than Ceph. LevelDB performs better on the backup,
as neither file system has to replicate.


\paragraph{Primary recovery.} After 30 s, we restart the file
system daemons on the primary, emulating the time for a machine reboot from
NVM. During this time, many file system operations occur on the backup
that need to be replayed on the primary. As soon as the primary is
back online, we cleanly close the database on the backup and restart
on the primary. Both \sys and Ceph allow applications to operate
during primary recovery, but performance is affected. \Sys detects
outdated files via epochs and reads their contents from the remote
cache replica upon access. Once read, the local copy is updated,
causing future reads to be local. LevelDB returns to full performance
938 ms after restarting it on the recovering primary. Ceph also
rebuilds the local OSD, but eagerly and in the background. Ceph takes
13.2 s before LevelDB serves its first operation due to contention
with OSD recovery and suffers another delay of 24.9 s on first
compaction, reaching full performance 43.4 s after recovery
start. \Sys recovers to full performance 46$\times$ faster than Ceph.

\paragraph{Fail-over to cold backup.} We measure cascaded LevelDB
fail-over time to an \sys replica with a cold cache. LevelDB serves
its first request on the cold backup 303 ms after failure detection,
but with higher latency due to SSD reads. LevelDB returns to full
performance after another 2.5 s. At this point, the entire dataset has
migrated back to cache.

\paragraph{Process fail-over.} For this benchmark, we simply kill
LevelDB. In this case, the failure is immediately detected by the
local OS and LevelDB is restarted. Ceph can reuse the shared kernel
buffer cache in DRAM, resulting in LevelDB restoring its database
after 1.63 s and returning to full performance after an additional
2.15 s, for an aggregate 3.78 s fail-over duration. With \sys, the DB
is restored in 0.71s, including recovery of the log of the failed
process and reacquisition of all leases. Full-performance operations
occur after an additional 0.16s, for an aggregate 0.87 s fail-over
time. \Sys recovers this case 4.34$\times$ faster than Ceph, showing
that process-local caches are not a hindrance to fast recovery.

\paragraph{OS fail-over.} NVM's performance allows for instant local
recovery of an OS failure, rather than requiring a backup replica. To
demonstrate, we run the primary in a virtual machine (VM). We kill the
primary VM, then immediately start a new VM from a snapshot stored in
NVM. The snapshot starts in 1.66 s. We restart \kernfs within the new
VM, which recovers the file system within 0.23 s. Finally, as in the
process fail-over experiment, LevelDB is restarted and resumes
database operations after another 0.68 s.  The aggregate fail-over
time is 2.57 s, comparable to \sys fail-over to a cold backup and
40$\times$ faster than Ceph's fail-over to a backup replica.




\subsection{Scalability}\label{sec:scalability}

\begin{figure}[t]
    \includegraphics[width=\columnwidth]{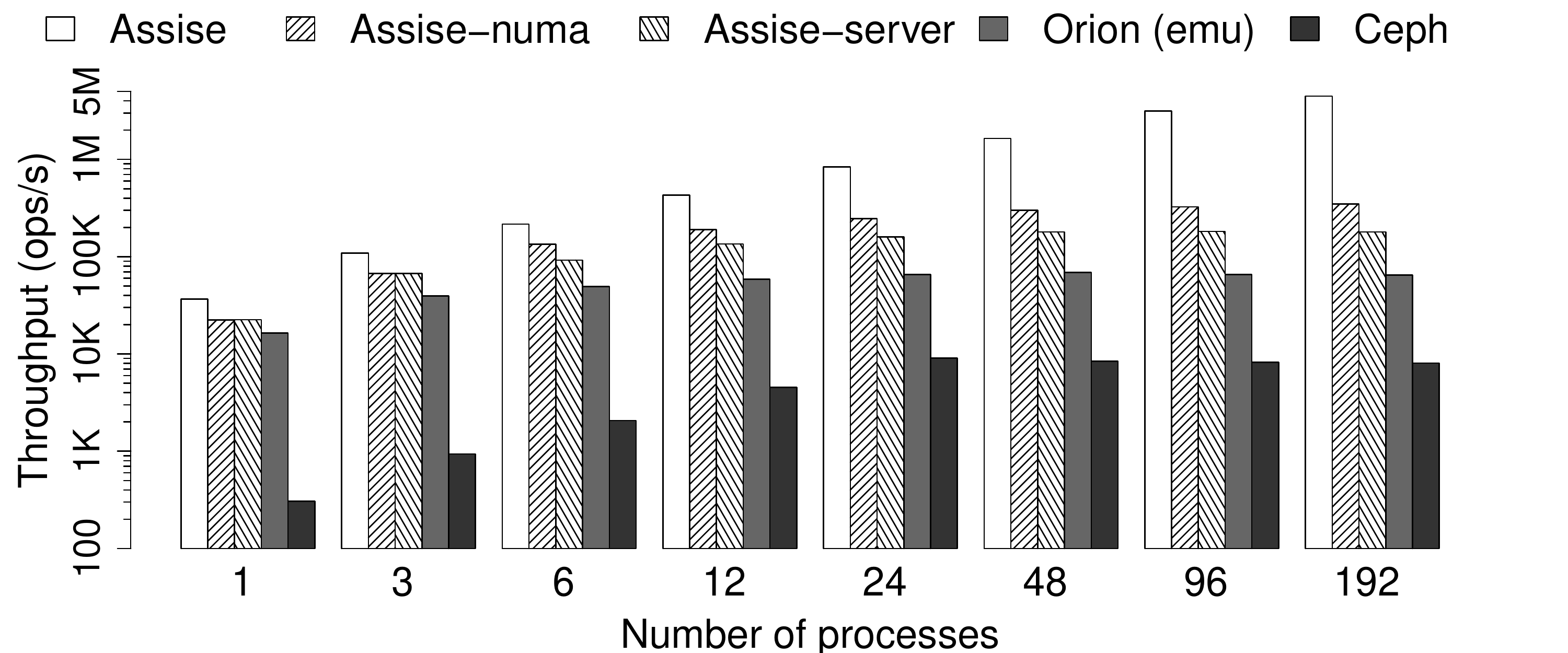}
  \caption{Scalability of atomic 4KB file operations. Log scale.}
  \label{fig:scalability}
\end{figure}

We evaluate \sys's scalability via 1) sharded file operations under
various levels of file system disaggregation, and 2) parallel email
delivery in Postfix~\cite{postfix}.

\subsubsection{Sharded file operations}

Processes in parallel create, write, and rename 4KB files with random
data in private directories. This benchmark uses 3 machines (6
sockets) and can scale throughput linearly with the number of
processes. To eliminate network bottlenecks to scalability, we turn
replication off. Figure~\ref{fig:scalability} presents average
throughput over 5 runs of an increasing number of processes, each
operating on 480K files, balanced over processor sockets. Ceph uses 3
sharded MDSes (1 per machine). However, MDS sharding had negligible
impact on Ceph's performance.

Ceph's disaggregated MDSes have high overhead for atomic operations,
as each client has to communicate with remote MDSes. This
disaggregated design prevents scalability beyond 8k ops/s. We emulate
Orion by restricting \ccnvm to use a single \kernfs lease manager. In
this case, data is stored on colocated NVM, but atomic operations
still use a remote lease manager. Orion has RDMA mechanisms that
simplify communicating with its MDS, but these mechanisms cannot be
used for operations that affect multiple inodes (\eg renames) and
Orion and \sys both use RDMA RPCs. While Orion operates in the kernel,
our emulation uses user-level RDMA, which is light-weight, and Orion
(emu) outperforms Ceph by 8$\times$.

To break down the benefit of local lease management in \sys, we
progressively shard it, first by server (\sys-server), then by socket
(\sys-numa), and finally by process (\sys). \Sys-server outperforms
Orion (emu) by 2.77$\times$ and \sys-numa improves throughput by
another 1.93$\times$. Finally, \sys scales linearly with the number of
processes until we hit NVM write bandwidth, improving throughput by
another 12.86$\times$. \sys outperforms Orion by 69$\times$ and Ceph
by 554$\times$ at scale.

\begin{figure}[t]
  \centering
  \includegraphics[width=\columnwidth]{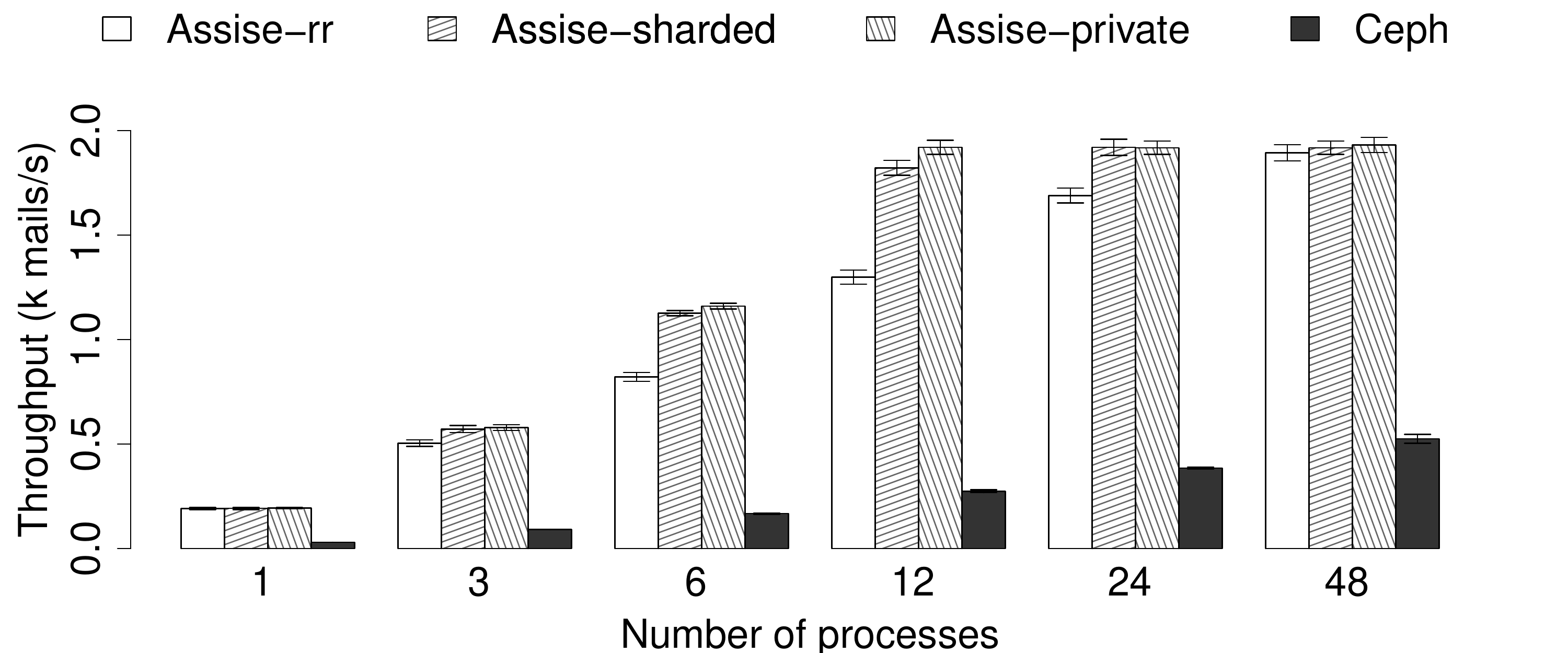}
  \caption{Postfix mail delivery throughput scalability.}
  \label{fig:postfix}
\end{figure}


\vspace{-1ex}
\subsubsection{Postfix}

We use the unmodified Postfix mail server to measure the performance
of parallel mail delivery. A load balancer machine forwards incoming
email from as many client machines as necessary to maximize throughput
to Postfix queue daemons running on 3 testbed machines, configured as
replicas. On each Postfix machine, a pool of delivery processes pull
email from the machine-local incoming mail queue and deliver it to
user Maildir directories on a cluster-shared distributed file
system. To ensure atomic mail delivery, a Postfix delivery process
writes each incoming email to a new file in a process-private
directory and then renames this file to the recipient's Maildir.

We send 80K emails from the Enron dataset~\cite{enron}, with each
email reaching an average of 4.5 recipients. This results in a total
of 360K email deliveries. Each email has an average size of 200 KB
(including attachments) and the dataset occupies 70 GB. We repeat each
experiment 3 times and report average mail delivery throughput and
standard deviation (error bars) in Figure~\ref{fig:postfix} over an
increasing number of delivery processes, balanced over machines. We
compare various \sys configurations and Ceph with 2 MDSes (1 MDS
performed similarly).


\paragraph{Round-robin.} In the first configuration (\sys-rr) the load
balancer uses a round-robin policy to send emails to mail queues. Due
to a lack of locality, mails delivered to the same Maildir often
require synchronization across machines, causing \ccnvm to frequently
delegate leases remotely and resulting in increased delivery
latencies. Despite this, \sys-rr is able to outperform Ceph by up to
5.6$\times$ at scale. Ceph cannot improve throughput much
further---even with 300 delivery processes, its throughput improves by
8\% versus 48 processes.

\paragraph{Sharded.} We shard Maildirs by Enron suborganization over
machines~\cite{grapevine}. 
The load balancer is configured to prefer the recipient's shard. For
mail messages with multiple recipients, it picks the shard with the
most receivers. In case of mail queue overload, the load balancer
sends mail to a random unloaded shard. Sharding users in this manner
provides up to 20\% better performance (\sys-sharded) due to the fact
that repeated deliveries to users of the same clique are likely to
occur on the same server, allowing \ccnvm to synchronize delivery
locally. At 15 processes, we are network-bound due to
replication. Sharding did not improve Ceph's performance.



\paragraph{Private directories.} We shard Maildirs by delivery
process, using process IDs for Maildir subdirectories, thereby
eliminating the need for synchronization (\sys-private).  This change
is not backward compatible with existing mail readers, but it is the
logical limit for sharding-based optimization.  \Sys-private scales
linearly until it is bottlenecked by network bandwidth, but
performance is similar to \sys-sharded. This shows that local
synchronization in \sys has minimal overhead. Ceph performance
continues to be gated by the MDS.

\paragraph{Summary.} Our evaluation demonstrates that localizing
consistency improves scalability for sharded workloads. Most
importantly, the results show \sys's ability to deliver almost the
same level of performance as private directories even if it has to
perform synchronization on shared directories.

\section{Conclusion}

We argue that NVM's unique characteristics require a redesign of
distributed file systems to cache and manage data on colocated
persistent memory.  We show how to leverage NVM colocation in the
design and implementation of \sys, a distributed file system that
provides low tail latency, high throughput, scalability, and high
availability with a strong consistency model.  \sys uses cache
replicas in NVM to minimize application recovery time and ensure data
availability, while leveraging a crash consistent file system cache
coherence layer (\ccnvm) to provide scalability. In comparing with
several state-of-the-art file systems, our results show that \sys
improves write latency up to 22$\times$, throughput up to 56$\times$,
fail-over time up to 103$\times$, and scalability up to 6$\times$
versus Ceph, while providing stronger consistency semantics.

\bibliographystyle{abbrv}
\bibliography{bibliography}

\appendix

\onecolumn
\begin{minipage}{\textwidth}
\centering
  \includegraphics[width=\textwidth]{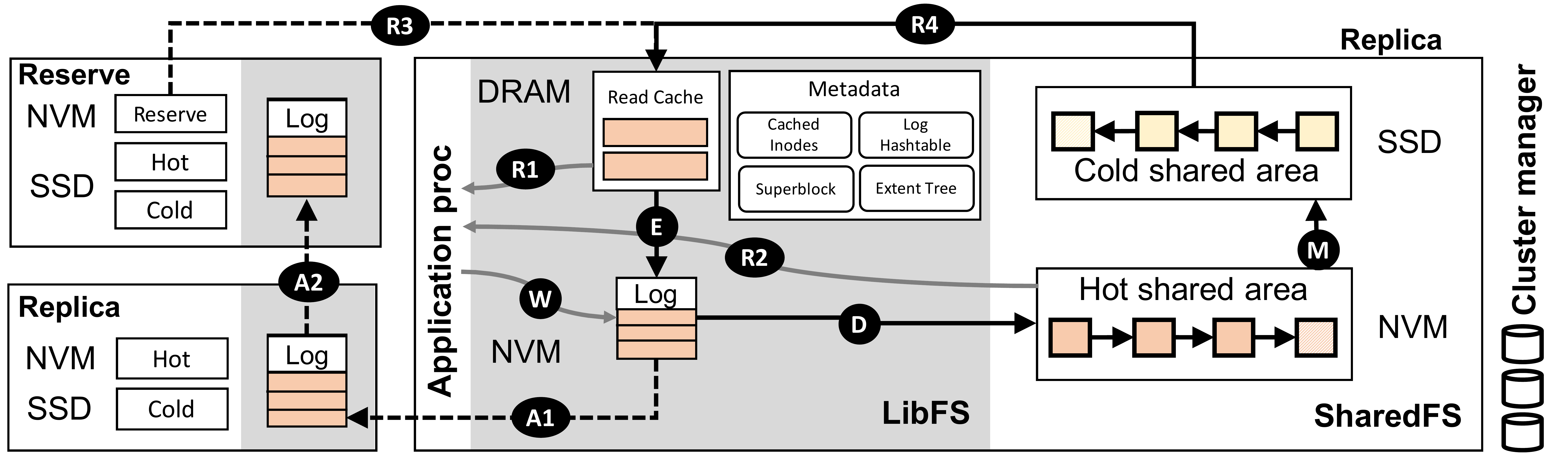}
  \captionof{figure}{\sys components and IO paths. Dashed line = RDMA operation, solid
    line = local operation. 
    Shaded areas are per process.
  }
  \label{fig:overview_appendix}
\end{minipage}
\begin{multicols}{2}
\section{IO Paths Appendix}\label{app:iopaths}

In this section of the appendix, we describe in detail the IO paths
introduced in Section~\ref{sec:iopaths}, including cold storage in
SSD, and how they integrate with Strata. We use some Strata
terminology in this section, which is defined in \cite{strata}.
Figure~\ref{fig:overview_appendix} illustrates the major IO paths in
\sys in detail.  Application processes using \sys run on a
configurable number of cache replicas.  For illustration, the figure
shows an example with 2 cache replicas and 1 reserve replica.



\subsection{Write Path}\label{app:write}

Writes in \sys involve three high-level mechanisms that operate on
different time scales: 1. To allow for persistence with low latency,
\libfs directly writes into a local, configurably sized, private
update log in NVM (\circled{W} in
Figure~\ref{fig:overview_appendix}). 2. The local update log is later
(on \texttt{fsync} when \emph{pessimistic}, on \texttt{dsync} when
\emph{optimistic}) chain-replicated by \libfs (\circled{A1},
\circled{A2}), which provides opportunities for saving network
bandwidth by \emph{coalescing}~\cite{strata} the log before
replication. 3. When the update log fills beyond a threshold, a
\emph{digest}~\cite{strata} of the log is initiated on every replica
(\circled{D}) to evict its contents. We describe replication and
digestion next.

\paragraph{Replication and crash consistency.}
When pessimistic, \texttt{fsync} forces immediate, synchronous
replication. The caller is blocked until all writes up to the
\texttt{fsync} have been replicated. Thus, all writes prior to an
\texttt{fsync} outlive node failures.

When optimistic, \texttt{fsync} is a no-op and \sys is free to delay
replication to coalesce more operations in the write log before
replication. In this case, \sys initiates replication on
\texttt{dsync} or upon digestion (see below).


After coalescing the local update log, its contents are written to the
\libfs private update log on the next replica along the replication
chain via RDMA writes (\circled{A1}). Finally, an RPC is sent to the
replica to initiate chain replication to the next replica
(\circled{A2}), and so on. The final replica in the chain sends an
acknowledgment back along the chain to indicate that the chain
completed successfully.

\paragraph{Digest/eviction.} When a process's private update log fills beyond a
threshold, \libfs replicates all log contents and then initiates a
digest on each replica along the replication chain via RPC
(\circled{A1}, \circled{A2}). Each replica checks log integrity and
potentially further coalesces them (\circled{D}). Each replica along
the chain digests in parallel and acknowledges when its digest
operation is finished.

\paragraph{Cold data migration.} Digests insert new data into \kernfs
\emph{hot shared areas}~\cite{strata} in NVM (the second-level cache),
migrating cold data out of these areas (\circled{M}). \Sys migration
is LRU based. Data migrates from private write log to hot, to optional
reserve, to cold shared area. For cache replicas, the hot shared area
resides in NVM; there is no reserve shared area, and a \emph{cold
  shared area} resides in SSD, which may be locally attached or
disaggregated (\eg via NVMe-over-Fabrics~\cite{nvmeof}). For the
reserve replica, the hot and cold shared areas both reside in SSD, and
there is a reserve shared area in NVM that is used to accelerate cold
reads (cf. \S\ref{app:read} and \S\ref{sec:reserve}).

\subsection{Read Path}\label{app:read}

\Sys allows different versions of the same file block to be
available in multiple storage layers simultaneously. The LRU data
migration mechanism guarantees that the latest version is always
available in the fastest media of the storage hierarchy. Upon a read,
\libfs 1. checks the process-private update log and DRAM read cache
(log hashtable and read cache in Figure~\ref{fig:overview_appendix}) for the
requested data block (\circled{R1}). If not found, \libfs 2. checks
the node-local hot shared area (\circled{R2}) via extent trees (cached
in process-local DRAM---extent tree in Figure~\ref{fig:overview_appendix}). If
the data was found in either of these areas, it is read locally. If
not found, \libfs 3. checks the reserve shared area on the reserve
replica (\circled{R3}), if it exists, and in parallel checks the cold shared
area on the local replica (\circled{R4}). If the data was found in the
reserve shared area, \libfs reads it remotely. Otherwise, it is read
locally.

\paragraph{Read cache management.} Recently read data is cached in
DRAM, except if it was read from local NVM, where DRAM caching does
not provide benefit. \Sys prefetches up to 256KB of data sequentially
when reading from cached media. The read cache caches 4KB blocks,
which is also the IO granularity of the SSD. Remote NVM reads can
happen at smaller granularity (\S\ref{sec:iopaths}). Filling the DRAM
cache with new data might necessitate evicting old data. In this case,
the data is written back from DRAM to NVM by \libfs to the local
update log (\circled{E}). The updated data migrates to the hot shared
area on digest. Finally, upon release of a lease, \libfs invalidates
corresponding cache entries.

\section{Sizing the Update Log}\label{app:log}

\Sys uses \libfs update logs in NVM per process and cache replica to
provide fast replicated IO with kernel-bypass. Update log space
requirements can impact scalability with many processes, depending on
available NVM space. In this section, we study the impact of the
\libfs update log size on \sys's performance and scalability. We also
explain how logs may be dynamically resized to adapt to workload
demands and available NVM capacity.

\subsection{Performance Sensitivity Analysis}\label{app:sensitivity}

\begin{figure}[H]
  \centering
  \includegraphics[width=\columnwidth]{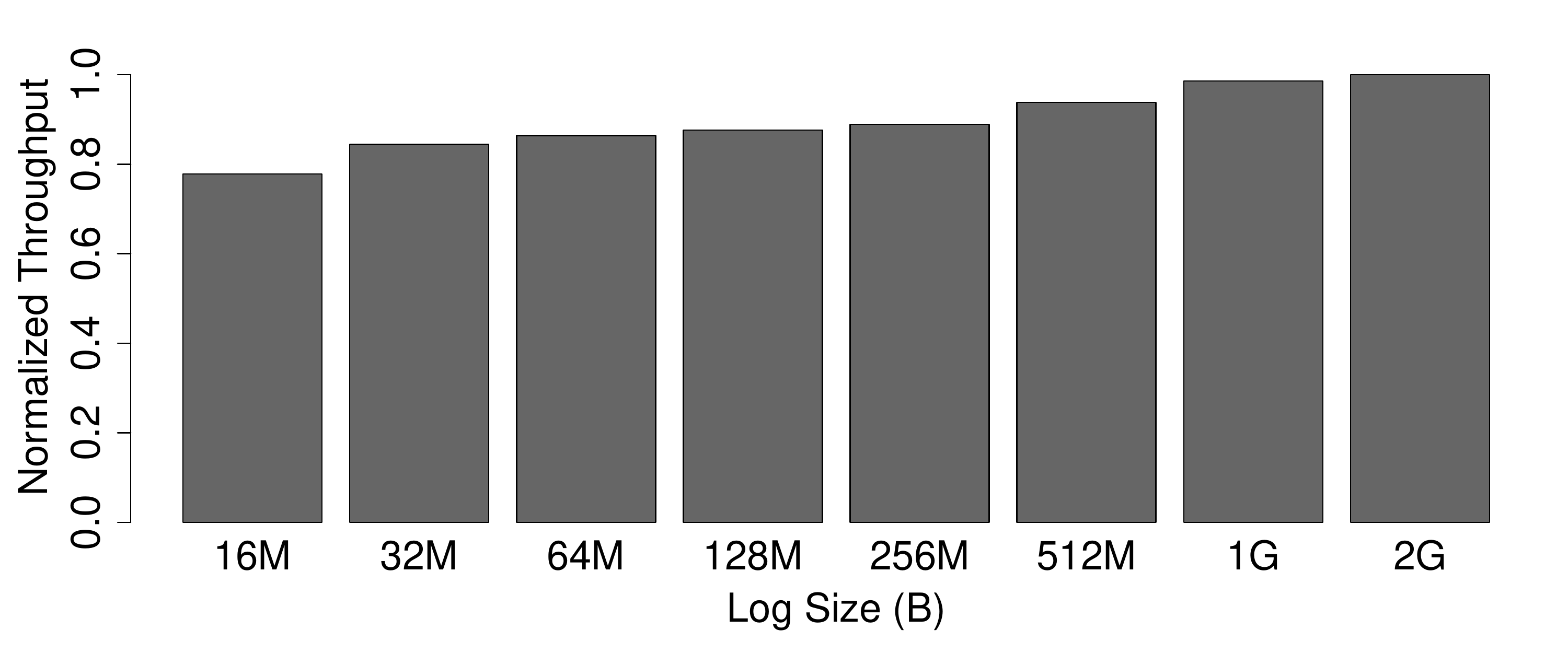}
  \caption{Throughput versus update log size, normalized to 2GB.}
  \label{fig:log_sensitivity}
\end{figure}

We perform a sensitvity analysis to evaluate the impact of LibFS
update log size on application performance. To do so, we run a
single-process microbenchmark that writes a 1 GB file sequentially at
4 KB IO granularity. We consider this as a worst case scenario for
smaller log sizes, since, in the absence of contention, processes can
quickly fill up their allocated log
space. Figure~\ref{fig:log_sensitivity} shows the normalized write
throughput at different log sizes. Throughput increases with log size,
saturating at 2GB, but the performance impact is small. Throughput
increases by only 22\% when using the largest (2GB) log size, which is
128$\times$ larger than the smallest (16MB). For workloads that share
data, we expect this gap to be smaller, as logs are evicted upon lease
handoff. With 6 TB of NVM per machine, \sys can scale to thousands of
processes even with 2GB update logs. At 16 MB, 100,000s processes can
be supported.

\subsection{Dynamic Log Sizing}\label{app:resizing}

It is possible to dynamically resize update logs to adapt to
momentarily available NVM capacity. \kernfs can resize logs upon
eviction/digestion, which is triggered whenever the log fills beyond a
threshold or upon explicit request by \libfs. \kernfs uses a two-phase
commit protocol to enforce identical log size across cache replicas
when resizing. The first phase asks all replicas to resize the
identified update log, which can be accepted or denied. If accepted by
all replicas, the resize operation is committed in the second phase or
aborted if at least one replica dissents (\eg on out of memory). When
accepting in the first phase, appropriate log space is simply reserved
and only allocated on commit.

The most significant overhead for log resizing is memory registration
for RDMA. It requires pinning the memory, assigning an identifier, and
mapping it in the RDMA NIC. However, this operation can be overlapped
with the digest itself. To help reduce the need for frequent resizing,
logs can be resized multiplicatively at first. After the log size
exceeds a certain threshold, its expansion can be limited to fixed
increments, similar to resizing approaches in prior work~\cite{nova}.



\section{Xfstests}\label{app:tests}

We tested \sys, NFS, and Ceph using xfstests~\cite{xfstests}. \Sys was
able to pass all 75 ``generic'' xfstests that are recommended for
NFS~\cite{nfstests}. Comparatively, NFSv4.2 and Ceph v14.2.1 pass only
71 and 69 of these tests, respectively.

\subsection{NFS}

NFS fails test cases 35, 423, 465, and 469.

\paragraph{Consistency (423, 465, 469).} To improve performance, NFS
clients cache file attributes using a delay-based heuristic. This
causes file attributes, such as the change time, to not be kept
immediately consistent when file attributes are changed (423). NFS
also does not provide consistency among direct IO (\verb+O_DIRECT+)
writes and buffered reads. Neither the data, nor the meta-data are
kept consistent (465). Similarly, consistency is not provided between
memory mapped IO and file meta-data changes. For example, a memory
mapped file that is later truncated does not update the memory map
immediately, continuing to expose the trunacted area to the memory map
(469).

\paragraph{Missing features (35).} Test 35 checks that a file or
directory that is open but has been overwritten will have no hard
links to it. This test fails for NFS files because NFS servers do not
track open files. Instead, NFS clients use a technique called
``silly-renaming'', where the open file is instead renamed to a hidden
file on the server before being overwritten, causing a hard link to
still exist.

\subsection{Ceph}

Ceph fails test cases 91, 213, 258, 263, 313, and 451.

\paragraph{Consistency (451, 313).} Ceph does not provide consistency
among an async direct (\verb+O_DIRECT+) write process and multiple
buffered read processes sharing a file~\cite{cephfs-posix}. Data read
by the readers does not match the data written by the writer. Ceph
does not update mtime after certain file operations, such as
truncation.

\paragraph{Crash consistency (258).} Negative ctimes underflow after
remount. For example, ctime -315593940 is changed to 3979380556 after
remounting.

\paragraph{Error mismatch (213).} The test tries to reserve (via
\verb+fallocate+) more space than the file system has. The expected
error is ``fallocate: No space left on device'' but Ceph reports
``fallocate: File too large''.

\paragraph{Missing features (91, 263).} The \verb+fallocate+ flags
\verb+FALLOC_FL_ZERO_RANGE+, \verb+FALLOC_FL_COLLAPSE_RANGE+, and
\verb+FALLOC_FL_INSERT_RANGE+ are not supported by Ceph. The tests are
thus incomplete, skipping some operations. Both tests are related to
direct IO and sub-block size buffered I/O. 91 does concurrent buffered
I/O.

\end{multicols}


\end{document}